\documentclass[11pt]{article}
\usepackage[english]{babel}
\usepackage{graphicx}
\usepackage[authoryear,round]{natbib}
\usepackage{rotating}
\usepackage[margin=20mm]{geometry}
\usepackage{hyperref}
\hypersetup{
    colorlinks = true,
    linkcolor = blue,
    anchorcolor = blue,
    citecolor = blue,
    filecolor = blue,
    urlcolor = blue,
	pdfauthor={some author},
	pdftitle={eye-catching title}
    }
\usepackage{amssymb,amsmath}
\usepackage{url}

\begin{document}

\title{Hunting the nature of the Seyfert 1 galaxy GRS~$1734-292$:\\
can it be a gamma-ray emitter?}
\author{Luigi Foschini\\
\small{Brera Astronomical Observatory}\\
\small{National Institute of Astrophysics (INAF)}\\
\small{Milano/Merate, Italy}\\
\small{email: \texttt{luigi.foschini@inaf.it}}}
\date{August 31, 2026}
\maketitle

\begin{abstract}
It has been proposed that the radio-weak Seyfert galaxy GRS~$1734-292$ is the counterpart of 3EG~J$1736-2908$ and of the nearby \emph{Fermi}/LAT source. I examine this identification, considering that the radio weakness could result from free-free absorption of a relativistic jet. An association with the EGRET detections is plausible, whereas the latest \emph{Fermi}/LAT localization no longer favors GRS~$1734-292$. New observations, especially at radio frequencies of tens of GHz, would help to settle the issue.
\end{abstract}

\section{Introduction}
\label{intro}
GRS~$1734-292$ ($z=0.02176$, \citealt{KOSS2022,MAGNO2025}) is a type-1 Seyfert active galactic nucleus (AGN) behind the Galactic plane, apparently close to the Galactic centre (at about $\sim 1.8^{\circ}$ from Sgr~A$^{*}$), a region of the sky affected by a significant hydrogen column ($N_{\rm H}^{\rm Gal}=6.51\cdot 10^{21}$~cm$^{-2}$, \citealt{HI4PI}). \cite{DICOCCO2004} proposed the association of GRS~$1734-292$ with the gamma-ray source 3EG~J$1736-2908$, although the radio weakness (only a few tens of mJy at 1.4 and 5 GHz) cast some doubt (see also \citealt{FOSCHINI2026} for a recent review on gamma-ray Seyfert, including this case). 

\begin{figure}[!h]
\begin{center}
\includegraphics[width=\textwidth]{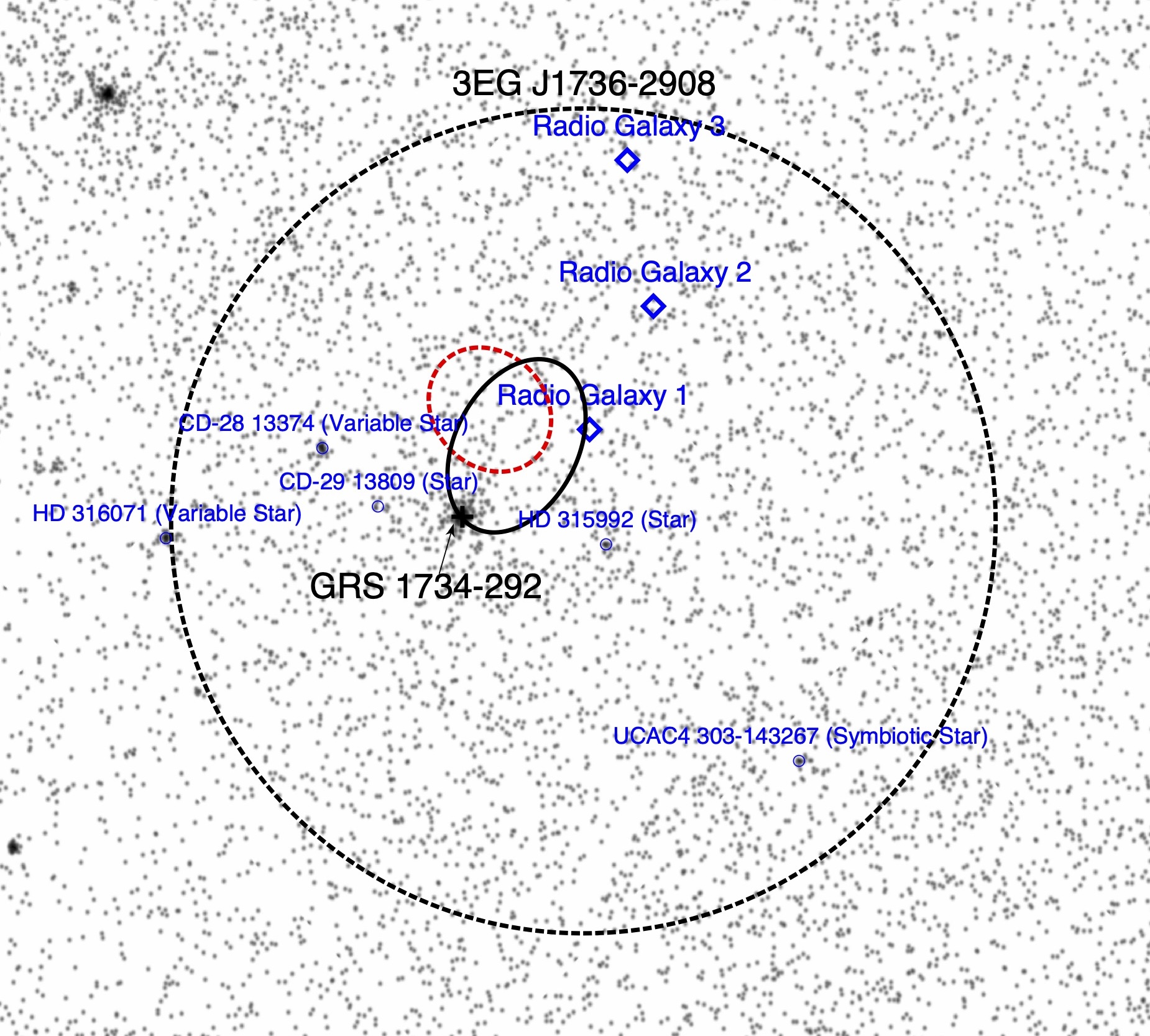}
\caption{\emph{SRG}/eROSITA sky map ($1-2$~keV) in the region around GRS~1734-292. Superimposed are the 95\% error contours of 3EG~J$1736-2908$ (dashed circle) and 4FGL~J$1737.1-2901$ (continuous ellipse). The red dashed ellipse it the 95\% error contour of FL16Y~J$1737.2-2858$. Radio Galaxies 1, 2, and 3 are from the radio survey by \cite{PATTIE2024}. eROSITA sky map has been generated by using the Aladin Sky Atlas Lite \citep{BAUMANN2022}.}
\label{fig:gammaxray}
\end{center}
\end{figure}

The Data Release 4 of the Fourth \emph{Fermi}-LAT Point Source Catalog (4FGL-DR4, \citealt{4FGLDR3,BALLET2023}) contains one source consistent with the EGRET source, which is 4FGL~J$1737.1-2901$. The 95\% error contour is an ellipse with semi-major axis $a_{95\%}=0^{\circ}.1387$, semi-minor axis $b_{95\%}=0^{\circ}.0922$, and position angle $62^{\circ}.57$, much smaller than the 95\% error contour of 3EG~J$1736-2908$ (radius $=0^{\circ}.62$), but still consistent with GRS~$1734-292$ (Fig.~\ref{fig:gammaxray}). Although the 4FGL-DR4 did not report any association with known sources, \cite{MICHIYAMA2024} took the association with GRS~1734-292 for granted, and reported a detection with Atacama Large Millimeter/submillimeter Array (ALMA) at $97.5$, $145$, and $225$ GHz, with variable flux. \cite{MICHIYAMA2024} suggested that it can be synchrotron emission from the corona or from a disk wind. The disk wind hypothesis was later favored by \cite{SAKAI2025} by modeling the spectral energy distribution (SED). Therefore, GRS~$1734-292$ might be a case similar to the type-1 Seyfert NGC~4151, whose gamma-ray emission might be due to ultrafast outflows \citep{PERETTI2025}. 

However, I would not completely discard the jet hypothesis. There is another important point to consider: \cite{LAHTEENMAKI2018} discovered intense (Jy-level) and erratic radio emission at $37$~GHz from narrow-line Seyfert 1 galaxies (NLS1s) that were formally classified as radio-quiet or even silent, thus indicating the presence of an intermittent relativistic jet (see also \citealt{OJHA2024} for a different interpretation). Radio follow-up showed that these sources are radio faint at frequencies $\lesssim 10$~GHz, while they are detected at higher frequencies with strong flux densities at Jy-level (\citealt{BERTON2020,JARVELA2021,JARVELA2024}). \cite{BERTON2020} suggested that their relativistic jets are likely to be absorbed at low frequencies via free-free absorption by a surrounding plasma sheet\footnote{The Galactic Center is characterized by the presence of vast regions rich in hot interstellar plasma, which can influence the radio emission coming from extragalactic sources behind it. \cite{ROY2013} studied the region within $\sim 2^{\circ}$ from the centre of the Milky Way, and found significant free-free Galactic absorption mostly within $\sim 0^{\circ}.6$ from the Galactic Centre. GRS~1734-292 is in the list of the studied sources with the name G$358.891+1.409$, but it is not found to be affected by free-free absorption of Galactic origin, being at $\sim 1.8^{\circ}$ from Sgr~A$^{*}$. Therefore, if the radio emission of GRS~1734-292 is affected by free-free absorption, it must be an intrinsic feature of the AGN.}, perhaps due to starburst activity -- common in NLS1s -- or shocks. Confirmation of this hypothesis came from \cite{ROMANO2023}, who carried out a \emph{Swift} monitoring of one of these sources -- SDSS J$164100.10+345452.7$ ($z=0.164$) -- between 2019 and 2021. The campaign revealed variable intrinsic absorption of the source in the X-ray energy band, which disappeared in the vicinity of the radio burst, while at the same time the X-ray photon index changed from $\Gamma\sim 1.9$ (with intrinsic absorption) to $\sim 0.7$ during the radio burst (no intrinsic absorption).

Perhaps, also the case of 1H~$0323+342$ ($z=0.063$) can display intriguing similarities with SDSS J$164100.10+345452.7$, showing spectral changes and variable absorption edges \citep{FOSCHINI2025}.

In the present work, I would like to reassess whether GRS~$1734-292$ might really be the counterpart of the gamma-ray source 3EG~J$1736-2908$/4FGL~J$1737.1-2901$, because its radio weakness might be due to free-free absorption at low frequencies. I will use a $\Lambda$CDM cosmology with $H_0=73.5$~km~s$^{-1}$~Mpc$^{-1}$ \citep{CASERTANO2026}. 


\section{Radio observations}
First, I searched for high-frequency radio observations ($\nu\gtrsim 10$~GHz), but I found only two in addition to the millimeter frequency detections by ALMA, still at mJy-level \citep{MICHIYAMA2024}. \cite{MARTI1998} reported Very Large Array (VLA) radio observations at 15~GHz, with a flux density of $8.7\pm 0.6$~mJy on April 10, 1997, while \cite{MAGNO2025} reported VLA radio observations at 22~GHz with a flux density of $7.50\pm 0.02$~mJy on Feb 27, 2022 (the exact date of observation was communicated to me via email by M. Magno). No other high-frequency radio observations were found. However, the intermittent jet hypothesis is not yet disfavored, because the duty cycle is rather erratic and only a long-term monitoring allowed the discovery of these outbursts \citep{LAHTEENMAKI2018}. 

\cite{PATTIE2024} performed a radio survey of the Galactic bulge at $1-2$~GHz and reported three radio galaxies within the 3EG~J$1736-2908$, although only one is consistent with 4FGL~J$1737.1-2901$. These radio sources are indicated in Fig.~\ref{fig:gammaxray} as RG1 (J$173635.9-290008$, peak flux density $\sim 42$~mJy), RG2 (J$173609.7-284901$, peak flux density $\sim 1.2$~mJy), and RG3 (J$173620.6-283551$, peak flux density $\sim 105$~mJy). However, none of these sources is detected at X-rays, making it unlikely to be good candidates for the gamma-ray source.

\section{X-ray data and analysis}
\label{datanalis}
I have found in the public archives 13 useful X-ray observations with GRS~$1734-292$ within the field-of-view (FOV) of the detectors, performed between 2008 and 2024. Details on specific satellites are given in the following Subsections. 

Other X-ray observations from different satellites were found in the literature and converted into the $0.3-10$~keV energy band (Table~\ref{tab:pastxray}). The reference model for the calculation of the unabsorbed extrapolated flux is (using the \texttt{xspec} syntax): \texttt{tbabs*ztbabs*zpo}, where $N_{\rm H}^{\rm Gal}=6.51\times 10^{21}$~cm$^{-2}$ \citep{HI4PI} is the Galactic absorption, $N_{\rm H}^{z}=1\times 10^{22}$~cm$^{-2}$ is the intrinsic absorption of the source given by the modeling of the \emph{Swift}, \emph{Chandra}, and \emph{XMM-Newton} data, $\Gamma=2$ (if no measurement was found), which is the standard assumption in these cases, and $z=0.02176$ \citep{KOSS2022}. This is also the starting model for the fit to the X-ray data. Absorption edges (\texttt{zedge}) are added if required at 99\% confidence level. 

For the conversion of the flux in Crab units into physical ones, I adopted the model given by \cite{KIRSCH2005}, that is $N_{\rm H}^{\rm Gal}=4.5\cdot 10^{21}$~cm$^{-2}$, $\Gamma=2.08$, $N=8.97$~ph~cm$^{-2}$~s$^{-1}$~keV$^{-1}$. 

\emph{SRG}/eROSITA catalog offers the flux in different energy bands, derived directly from the count rates by using an energy conversion factor \citep{MERLONI2024}: I have selected the $4-10$~keV band, because it is less affected by the absorption, and extrapolated to the $0.3-10$~keV band.

\subsection{\emph{Swift}/XRT}
I have found 69 \emph{Swift} observations within 10 arcminutes from the coordinates of GRS~$1734-292$, executed from September 6, 2012 to August 28, 2024. Exposures on the X-Ray Telescope (XRT) range from 42~s to $\sim 6.2$~ks. Given the high X-ray flux ($\sim 10^{-10}$~erg~cm$^{-2}$~s$^{-1}$), even very short exposures of tens of seconds can give a rough idea of the flux. However, since I am interested in finding spectral variations and/or absorption edges, these snapshots are not suitable. After having removed these observations, only six observations remained, with exposures starting from $\sim 470$~s. 

For the reduction and analysis of XRT data, I used \texttt{HEASoft v. 6.36} with \texttt{CALDB} updated on March 23, 2026, and followed the standard procedures described in \cite{CAPALBI2005}. All XRT observations were done in photon counting mode: given the high flux, I adopted an annular extraction region (inner radius $9.4''$, outer radius $40''$) to minimize the effects of pile-up, as suggested by \cite{EVANS2014}. In the case of ObsID $00015399001$, I have applied an internal radius of $14''$, because of some residual feature in the point-spread function (PSF) profile. The best-fit results are displayed in Table~\ref{tab:swift}.

\subsection{\emph{Chandra}/ACIS}
\emph{Chandra} archive contains six observations with GRS~$1734-292$ in the FOV, all with the Advanced CCD Imaging Spectrometer for Imaging (ACIS-I, front illuminated), with exposures ranging from $\sim 1.9$ to $\sim 5.3$~ks. For the reduction and analysis, I used \texttt{CIAO v. 4.18} with \texttt{CALDB v. 4.12.3}. The source is placed always on the border of the chip, to minimize the effects of pile-up. To further strengthen the correction of the pile-up, I selected events with grade 0, status 0, from an annular extraction region with inner radius of $6''$ and outer radius of $12''$. I followed the standard procedures described in the \href{https://cxc.cfa.harvard.edu/ciao/threads/all.html}{Science Threads}. The obtained spectra were fit in the $0.3-7.0$~keV energy band to take into account the calibration status, while the measured flux was then extrapolated to the $0.3-10$~keV band.  The ObsID $24025$ was divided into two bins of $\sim 2.2$ and $3.1$~ks, respectively, to search for spectral changes. The best-fit results are displayed in Table~\ref{tab:chandra}.

\subsection{\emph{XMM-Newton}/EPIC}
Only one observation is available in the \emph{XMM-Newton} public archive (obsID 0550451501, Feb 26, 2009, 13:37:28 UTC), with $\sim 17.6$~ks exposure. The count rates on the three detectors (MOS1, MOS2, pn) of the European Photon Imaging Camera (EPIC) exceeded the pile-up limit, so that I adopted annular extraction region (inner radius $20''$, outer radius $40''$) and selected pattern 0 events to minimize the effects of pile-up. To reprocess and analyze EPIC data, I used the \texttt{Science Analysis Software v. 22.1.0} and the calibration files (\texttt{CCF}) updated on March 23, 2026. Then, I followed the standard procedures described in the \href{https://www.cosmos.esa.int/web/xmm-newton/sas-threads}{Science Threads}. The observation was divided into five bins with hourly elapsed times. The best-fit results are displayed in Table~\ref{tab:xmm}.

\begin{sidewaystable}
\renewcommand{\arraystretch}{1.2}
\caption{Past X-ray observations from literature. Column explanation: (1) Name of the satellite and the instrument; (2) Observation date [UTC]; (3) Energy range [keV]; (4) Photon Index [(*) fixed when no measurement was available]; (5) Original observed flux [different units]; (6) Bibliographic reference; (7) Unabsorbed extrapolated flux in the $0.3-10$~keV energy band [$10^{-11}$~erg~cm$^{-2}$~s$^{-1}$]. }
\begin{tabular}{lcccclc}
\hline
Satellite & Date & Energy & $\Gamma$ & Original Flux & Reference & Flux\\
(1)			& (2) & (3) & (4) & (5) & (6) & (7)\\
\hline
\emph{GRANAT}/ART-P & Sep 7 - Oct 18, 1990 & $4-20$ & $2.0\pm 0.1$ & $(6.2\pm0.8)\cdot 10^{-3}$ & \cite{PAVLINSKY1994} & $18.3$\\
{} & {} & {} & {} & [ph~cm$^{-2}$~s$^{-1}$] & {} & {}\\
\emph{GRANAT}/ART-P & Oct 6, 1990 & $3-12$ & $2.0\pm 0.1$ & $3.4\pm 0.4$ & \cite{SUNYAEV1990} & $15.9$\\
{} & {} & {} & {} & [mCrab] & \cite{BARRET1996} & {}\\
\emph{ROSAT}/PSPC & Feb 29, 1992 & $0.1-2.4$ & $2.0^{*}$ & $(3\pm1)\cdot 10^{-2}$ & \href{https://heasarc.gsfc.nasa.gov/W3Browse/rosat/rospspc.html}{ROSPSPC/2RXP} & $0.764$\\
{} & 15:16:57 & {} & {} & [counts/s] & {} & {}\\
\emph{ROSAT}/PSPC & Feb 29, 1992 & $0.1-2.4$ & $2.0^{*}$ & $(8\pm2)\cdot 10^{-2}$ & \href{https://heasarc.gsfc.nasa.gov/W3Browse/rosat/rospspc.html}{ROSPSPC/2RXP}  & $1.63$\\
{} & 21:38:57 & {} & {} & [counts/s] & {} & {}\\
\emph{GRANAT}/SIGMA & Sep 15-17, 1992 & $40-400$ & $0.6\pm 0.4$ & $36$ & \cite{CHURAZOV1992} & 0.482\\
{} & {} & {} & {} & [mCrab] & {} & {}\\ 
\emph{ROSAT}/HRI & Mar 28-31, 1995 & $0.1-2.4$ & $2.0^{*}$ & $(6.6\pm0.8)\cdot 10^{-3}$ & \cite{BARRET1996} & $0.371$\\
{} & {} & {} & {} & [counts/s] & {} & {}\\
\emph{ASCA}/GIS & Sep 8, 1998 & $0.7-10$ & $1.41_{-0.50}^{+0.52}$ & $4.3$ & \cite{SAKANO2002} & $6.18$\\
{} & {} & {} & {} & [$10^{-11}$~erg~cm$^{-2}$~s$^{-1}$] & {} & {}\\
\emph{ASCA}/GIS & Mar 12, 1999 & $0.7-10$ & $1.51_{-0.19}^{+0.20}$ & $3.1$ & \cite{SAKANO2002} & $5.06$\\
{} & {} & {} & {} & [$10^{-11}$~erg~cm$^{-2}$~s$^{-1}$] & {} & {}\\
\emph{SRG}/eROSITA & Mar 27, 2020 & $4-10$ & $2.0^{*}$ & $3.25$ & \cite{MERLONI2024} & $12.8$\\
{} & {} & {} & {} & [$10^{-11}$~mW~m$^{-2}$] & {} & {}\\
\hline
\end{tabular}
\label{tab:pastxray}
\end{sidewaystable}

\begin{sidewaystable}
\renewcommand{\arraystretch}{1.5}
\caption{\emph{Swift} observations. Column explanation: (1) Observation Identifier; (2) Observation date [UTC]; (3) Exposure [ks]; (4) Intrinsic absorption [$10^{22}$~cm$^{-2}$; model \texttt{ztbabs}]; (5) power-law photon index [model \texttt{zpow}]; (6) edge energy [keV; model \texttt{zedge}]; (7) edge optical depth [model \texttt{zedge}]; (8) edge energy [keV; model \texttt{zedge}]; (9) edge optical depth [model \texttt{zedge}]; (10) unabsorbed flux [$0.3-10$~keV, $10^{-11}$~erg~cm$^{-2}$~s$^{-1}$]; (11) used statistic $\chi^2$ or likelihood (denoted with an asterisk); (12) degrees of freedom.}
\centering
\begin{tabular}{cccccccccccc}
\hline
ObsID & Date & Exp &  $N_{\rm H}^{z}$ & $\Gamma$ & $E_{1}$ & $\tau_1$ & $E_2$ & $\tau_2$ & Flux & Stat & dof\\
(1)	  & (2) & (3) & (4) & (5) & (6) & (7) & (8) & (9) & (10) & (11) & (12)\\
\hline
00032360003 & Sep 06, 2012, 21:43:07 & 0.5 & $1.81_{-1.02}^{+1.21}$ & $1.91_{-0.69}^{+0.75}$ & {} & {} & {} & {} & 17.2 & $104.73^{*}$ & $102$\\ 
00080187002 & Sep 16, 2014, 13:24:31 & 6.2 & $1.04_{-0.23}^{+0.25}$ & $1.57_{-0.15}^{+0.16}$ & {} & {} & {} & {} & 12.6 & $94.60$ & $89$\\ 
00088136001 & May 28, 2018, 20:24:24 & 2.0 & $1.85_{-0.76}^{+0.87}$ & $1.70_{-0.45}^{+0.48}$ & {} & {} & {} & {} & 9.67 & $144.31^{*}$ & $218$\\
00015399001 & Oct 30, 2022, 22:07:39 & 1.0 & $2.41_{-1.23}^{+1.51}$ & $1.79_{-0.86}^{+0.94}$ & $1.92_{-0.09}^{+0.13}$ & $2.1_{-1.2}^{+1.4}$ & $4.25_{-0.17}^{+0.23}$ & $1.3_{-0.8}^{+0.9}$ & $21.6$ & $113.51^{*}$ & $125$\\ 
00015942001 & Apr 04, 2023, 21:46:39 & 1.6 & $1.03_{-0.65}^{+0.79}$ & $1.24_{-0.48}^{+0.52}$ & {} & {} & {} & {} & 6.82 & $122.99^{*}$ & $176$\\
00015399002 & Aug 28, 2024, 01:33:54 & 1.7 & $0.82_{-0.73}^{+0.84}$ & $1.59_{-0.46}^{+0.51}$ & {} & {} & {} & {} & 11.4 & $20.76$ & $19$\\ 
\hline
\end{tabular}
\label{tab:swift}
\end{sidewaystable}

\begin{sidewaystable}
\renewcommand{\arraystretch}{1.5}
\caption{\emph{Chandra} observations. Column explanation: (1) Observation Identifier; (2) Observation Date [UTC]; (3) Exposure [ks]; (4) Intrinsic absorption [$10^{22}$~cm$^{-2}$; model \texttt{ztbabs}]; (5) power-law photon index [model \texttt{zpow}]; (6) edge energy [keV; model \texttt{zedge}]; (7) edge optical depth [model \texttt{zedge}]; (8) edge energy [keV; model \texttt{zedge}]; (9) edge optical depth [model \texttt{zedge}]; (10) unabsorbed flux [$0.3-10$~keV, $10^{-11}$~erg~cm$^{-2}$~s$^{-1}$]; (11) statistic $\chi^2$; (12) degrees of freedom. The ObsID 24025 has been divided into two bins with similar elapsed time ($\sim 3.3$~ks), but different exposures ($\sim 2.2$ and $3.1$~ks, respectively).}
\centering
\begin{tabular}{cccccccccccc}
\hline
ObsID & Date & Exp & $N_{\rm H}^{z}$ & $\Gamma$ & $E_{1}$ & $\tau_1$ & $E_2$ & $\tau_2$ & Flux & Stat & dof\\
(1)	& (2) & (3) & (4) & (5) & (6) & (7) & (8) & (9) & (10) & (11) & (12)\\
\hline
8702  & May 18, 2008, 04:24:17 & 2.2 & $1.74_{-0.45}^{+0.50}$ & $2.23_{-0.33}^{+0.35}$ & {} & {} & {} & {} & 4.93 & $94.35$ & $88$\\
8691  & May 18, 2008, 13:28:08 & 2.1 & $1.67_{-0.28}^{+0.31}$ & $1.99_{-0.20}^{+0.21}$ & {} & {} & {} & {} & 6.78 & $118.46$ & $107$\\ 
18656 & Jul 31, 2016, 00:50:33 & 2.0 & $1.73_{-0.37}^{+0.41}$ & $2.06_{-0.23}^{+0.24}$ & {} & {} & {} & {} & 6.27 & $94.35$ & $88$\\
18657 & Jul 31, 2016, 02:08:13 & 1.9 & $1.91_{-0.39}^{+0.43}$ & $2.02_{-0.23}^{+0.25}$ & {} & {} & {} & {} & 8.15 & $86.96$ & $94$\\
18658 & Jul 31, 2016, 02:53:03 & 1.9 & $2.08_{-0.49}^{+0.54}$ & $2.50_{-0.38}^{+0.40}$ & $1.68_{-0.08}^{+0.05}$ & $0.74_{-0.36}^{+0.39}$ & {} & {} & 16.5 & $53.87$ & $54$\\
24025A & Oct 02, 2022, 12:02:20 & 2.2 & $1.96_{-0.44}^{+0.47}$ & $2.69_{-0.33}^{+0.34}$ & $1.72_{-0.08}^{+0.06}$ & $0.51_{-0.27}^{+0.29}$ & {} & {} & 13.9 & $99.24$ & $90$\\
24025B & Oct 02, 2022, 12:57:21 & 3.1 & $2.32_{-0.42}^{+0.44}$ & $3.08_{-0.31}^{+0.32}$ & $1.60\pm 0.05$ & $0.75_{-0.29}^{+0.31}$ & $2.62_{-0.10}^{+0.07}$ & $0.41_{-0.20}^{+0.21}$ & 28.5 & $120.34$ & $112$\\
\hline
\end{tabular}
\label{tab:chandra}
\end{sidewaystable}

\begin{sidewaystable}
\renewcommand{\arraystretch}{1.5}
\caption{\emph{XMM-Newton} observation. Column explanation: (1) Hourly bin EPIC/pn exposure [ks]; (2) Intrinsic absorption [$10^{22}$~cm$^{-2}$; model \texttt{ztbabs}]; (3) power-law photon index [model \texttt{zpow}]; (4) edge energy [keV; model \texttt{zedge}]; (5) edge optical depth [model \texttt{zedge}]; (6) edge energy [keV; model \texttt{zedge}]; (7) edge optical depth [model \texttt{zedge}]; (8) unabsorbed flux [$0.3-10$~keV, $10^{-11}$~erg~cm$^{-2}$~s$^{-1}$]; (9) statistic $\chi^2$; (10) degrees of freedom.}
\centering
\begin{tabular}{cccccccccc}
\hline
H-Bin Exp & $N_{\rm H}^{z}$ & $\Gamma$ & $E_{1}$ & $\tau_1$ & $E_2$ & $\tau_2$ & Flux & Stat & dof\\
(1)			& (2) & (3) & (4) & (5) & (6) & (7) & (8) & (9) & (10)\\
\hline
1.9  & $0.96_{-0.17}^{+0.18}$ & $1.64_{-0.12}^{+0.13}$ & {} & {} & {} & {} & 9.64 & $110.39$ & $107$\\ 
3.1  & $1.00_{-0.15}^{+0.17}$ & $1.58\pm 0.11$ & {} & {} & {} & {} & 9.42 & $123.35$ & $129$\\
3.1  & $0.94_{-0.15}^{+0.16}$ & $1.57\pm 0.11$ & {} & {} & {} & {} & 9.14 & $159.21$ & $127$\\ 
3.1  & $0.98_{-0.16}^{+0.17}$ & $1.54\pm 0.11$ & {} & {} & {} & {} & 9.83 & $126.56$ & $127$\\
2.6  & $1.01_{-0.18}^{+0.19}$ & $1.53\pm 0.12$ & {} & {} & {} & {} & 9.61 & $126.69$ & $111$\\  
\hline
\end{tabular}
\label{tab:xmm}
\end{sidewaystable}

\section{Gamma-ray data and analysis}
\label{datagamma}

\subsection{\emph{CGRO}/EGRET}
\begin{figure}[h]
\begin{center}
\includegraphics[scale=0.1]{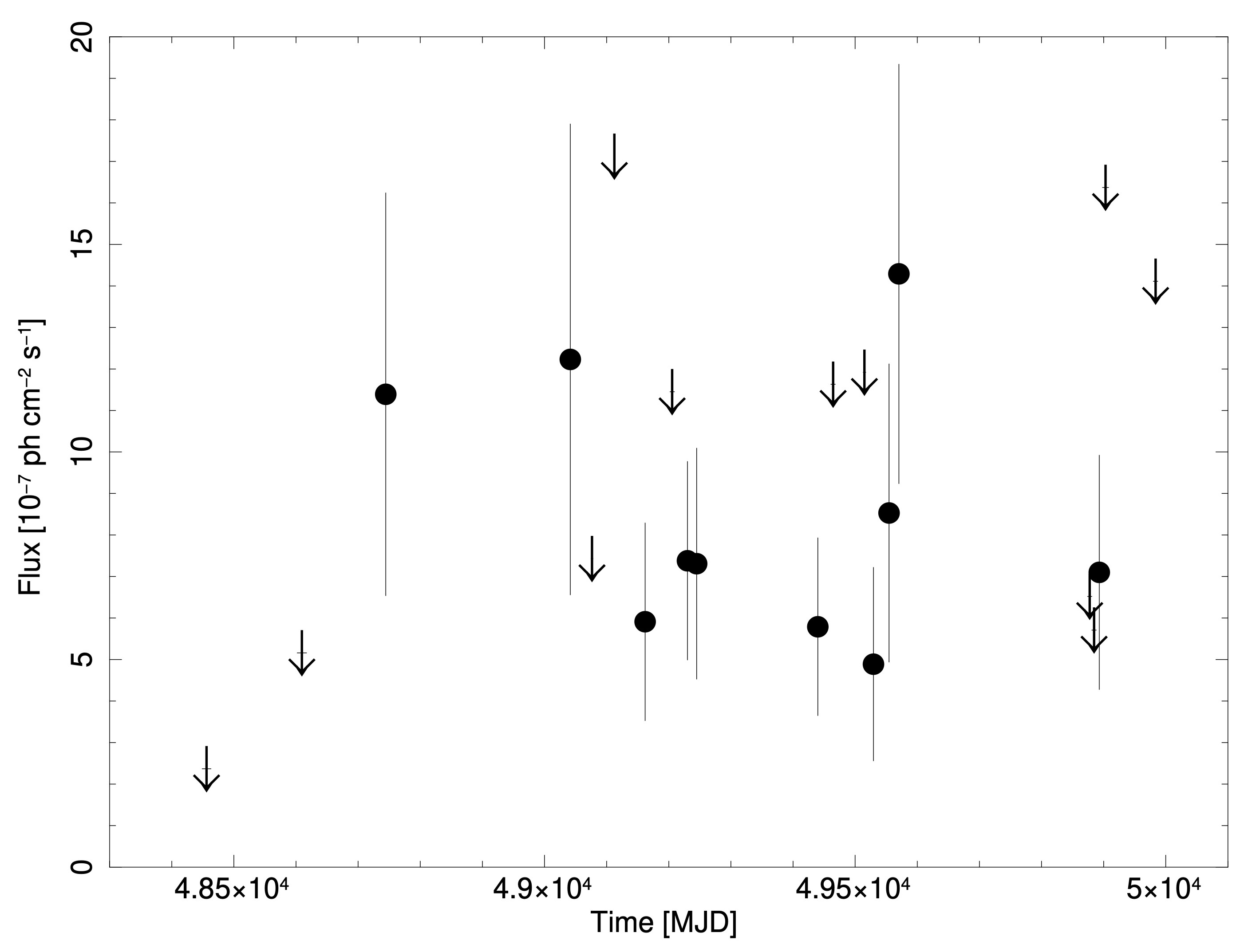}
\includegraphics[scale=0.1]{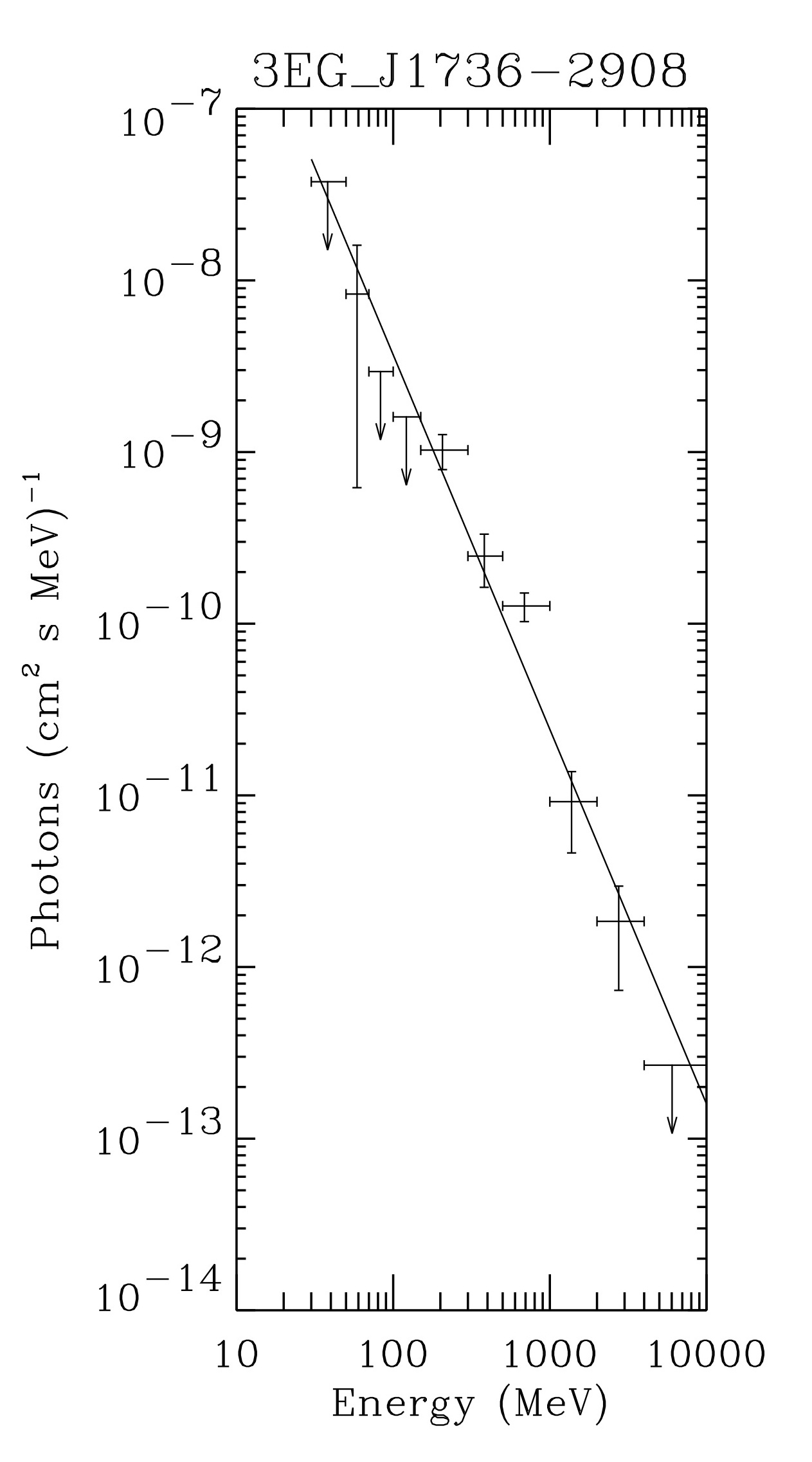}
\caption{(\emph{left panel}) Light curve of 3EG~J$1736-2908$. Time bins are the viewing periods, spanning from a few days to a few weeks. The first VP is $5.0$ (July 12-27, 1991), while the last one is $429.0$ (September 20-28, 1995). Upper limits are at 95\% confidence level. (\emph{right panel}) Integrated spectrum. VP data were taken from \cite{HARTMAN1999}, while the figure of the spectrum has been downloaded from the \emph{CGRO}/EGRET \href{https://heasarc.gsfc.nasa.gov/FTP/compton/data/egret/}{public archive} at HEASARC.}
\label{fig:egret}
\end{center}
\end{figure}

The Energetic Gamma Ray Experiment Telescope (EGRET) onboard the \emph{Compton Gamma-Ray Observatory} (CGRO) was a spark chamber working in the $0.1-20$~GeV energy band. Its third and last catalog contained 271 gamma-ray sources, with the flux measurements and upper limits divided by viewing periods (VPs) ranging from a few days to a few weeks \citep{HARTMAN1999}. I have taken the available data of 3EG~J$1736-2908$ divided per viewing periods. It is worth noting that \cite{HARTMAN1999} considered a test statistic (TS) greater than 4 already a detection, although it is equivalent to $\sim 2\sigma$. The resulting light curve and spectrum are displayed in Fig.~\ref{fig:egret}.

\begin{figure}[!ht]
\begin{center}
\includegraphics[scale=0.32]{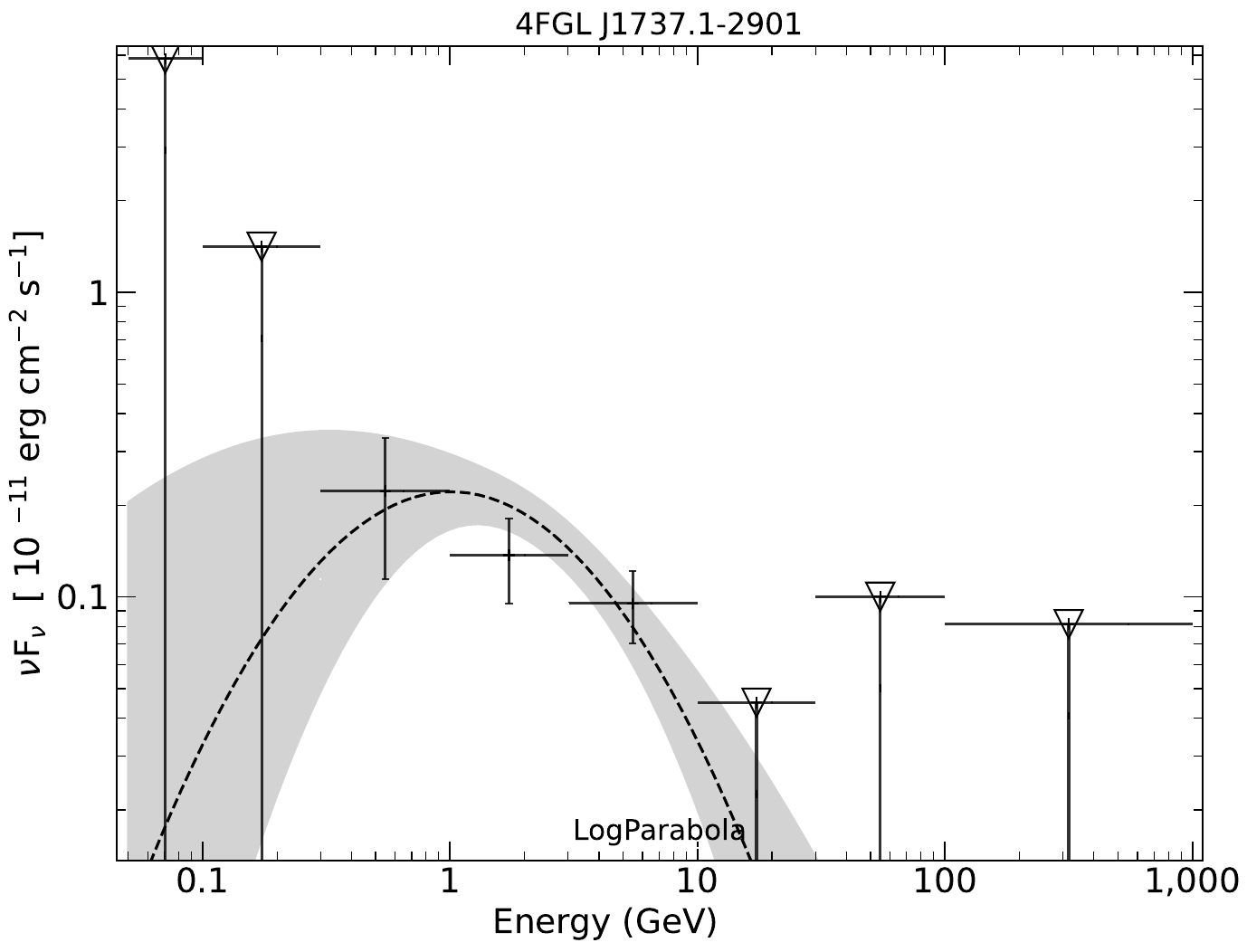}
\includegraphics[scale=0.32]{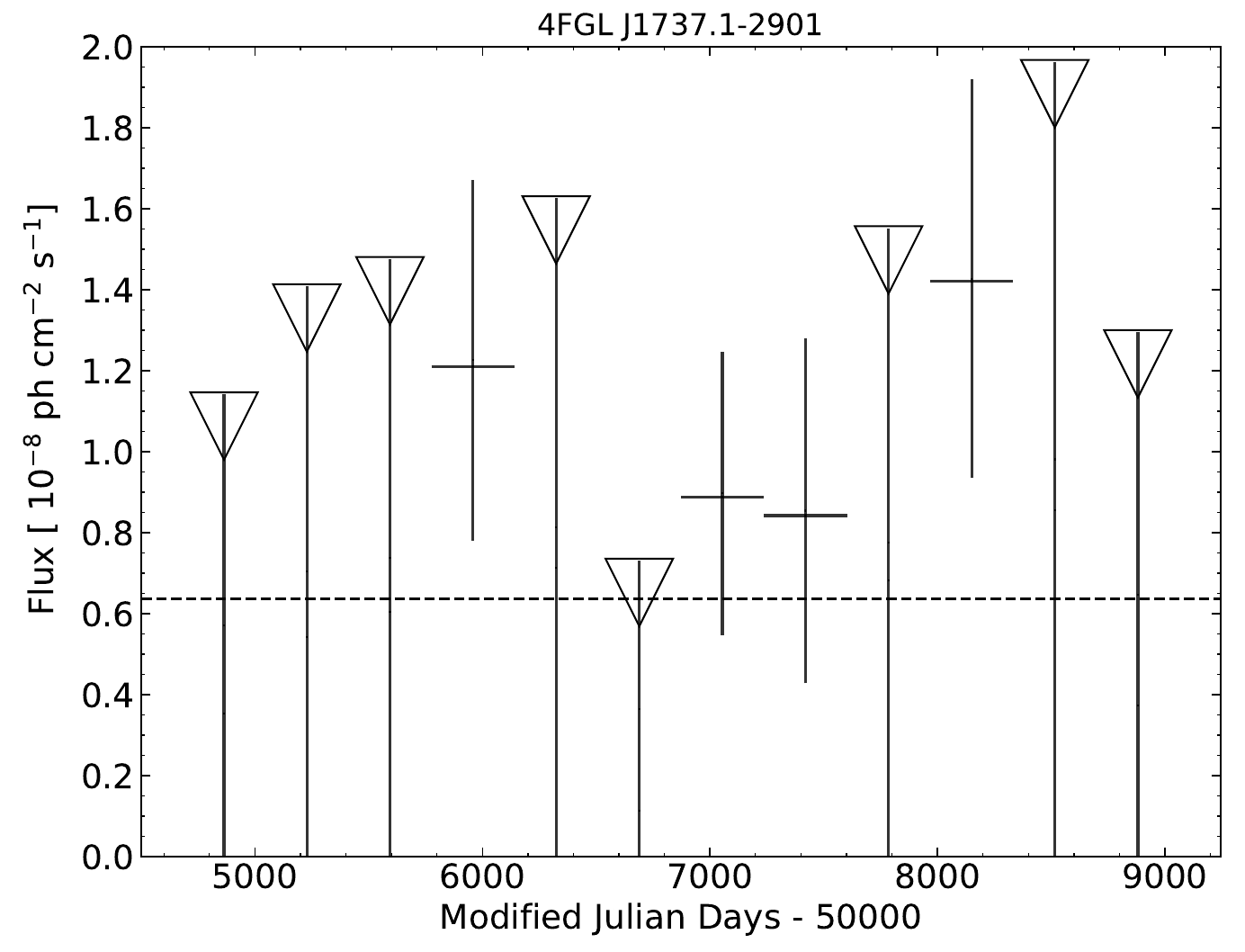}\\
\includegraphics[scale=0.085]{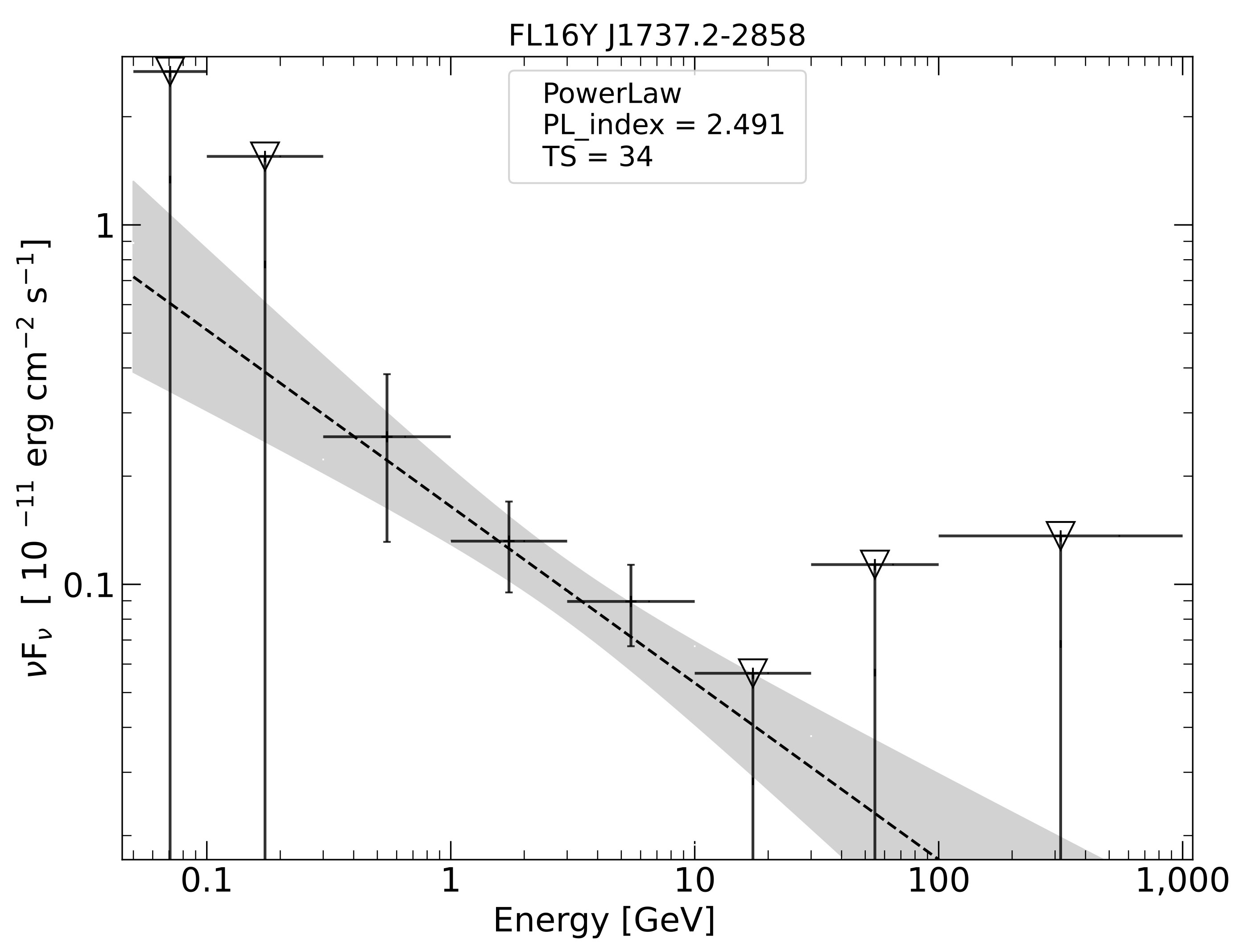}
\includegraphics[scale=0.079]{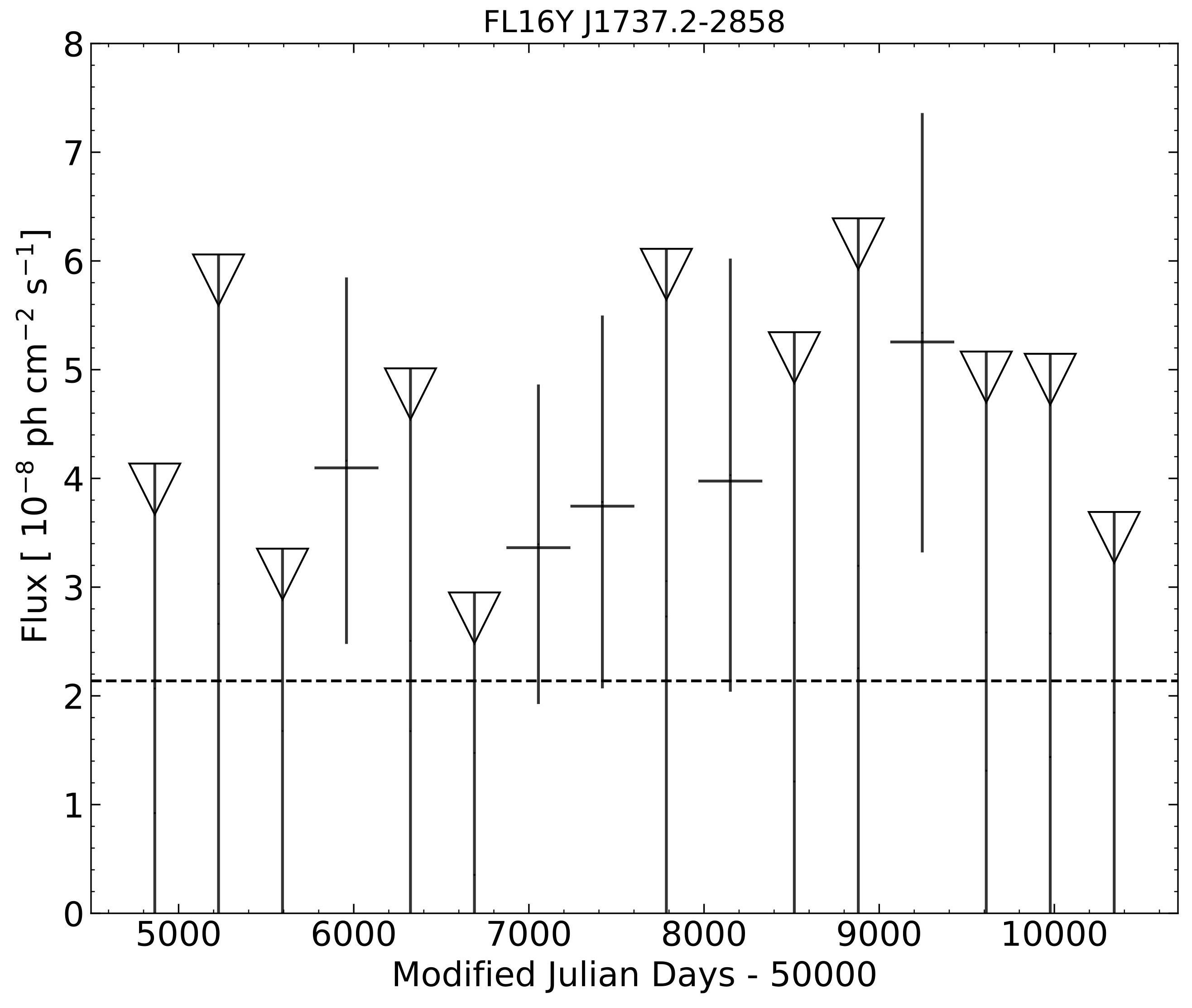}
\caption{Data products of the 4FGL and FL16Y \citep{4FGLDR3} from HEASARC. Top panels: spectrum (\emph{left}) and light curve (\emph{right}) of the gamma-ray source 4FGL~J$1737.1-2901$ \citep{BALLET2023}. Bottom panels: spectrum (\emph{left}) and light curve (\emph{right}) of the gamma-ray source FL16Y~J$1737.2-2858$ \citep{BALLET2026}.}
\label{fig:lat}
\end{center}
\end{figure}

\subsection{\emph{Fermi}/LAT}
I have already written in the Sect.~\ref{intro} about the gamma-ray source 4FGL~J$1737.1-2901$. However, the \emph{Fermi}-LAT Collaboration recently released the 16-years catalog (FL16Y, \citealt{4FGLDR3,BALLET2026}). The gamma-ray source is now FL16Y~J$1737.2-2858$, and its 95\% error ellipse has semimajor axis $a_{95\%}=0^{\circ}.1002$, semiminor axis $b_{95\%}=0^{\circ}.0846$, and position angle $-47^{\circ}.28$ (Fig.~\ref{fig:gammaxray}). GRS~$1734-292$ is now outside the 95\% error ellipse, but still consistent if we extend the error ellipse to 99\% confidence level. It is worth reminding that the gamma-ray source is close to the Galactic centre (less than $2^{\circ}$), in a region dominated by the Galactic diffuse emission. The analysis flags of 4FGL~J$1737.1-2901$ and FL16Y~J$1737.2-2858$ are 8192 and 8196, respectively, indicating a strong contamination with background at low energy \citep{BALLET2026}. 

As written above, if GRS$1734-292$ is a case similar to NGC~$4151$, the gamma-ray emission from winds could be concentrated at low energies, right in the band most contaminated by diffuse galactic emission ($\lesssim 1$~GeV). It is also necessary to take into account that NGC~$4151$ is closer to the Earth: $\sim 13.6$~Mpc to be compared with the $\sim 90.2$~Mpc of GRS~$1734-292$, which means that, given the same luminosity, the observed flux of GRS~$1734-292$ is about 44 times lower than that of NGC~4151.

The most recent catalogs, 4FGL and FL16Y \citep{4FGLDR3}, reported two slightly different gamma-ray sources: 4FGL~J$1737.1-2901$, consistent with GRS~$1734-292$, and FL16Y~J$1737.2-2858$, which is not consistent. Fig.~\ref{fig:lat} displays the average spectra and light curves of the two gamma-ray sources. There are some slight differences: the average spectrum of 4FGL~J$1737.1-2901$ is a log-parabola with $\Gamma=2.54\pm 0.24$ and $\beta=0.21\pm 0.14$, while in the case of FL16Y~J$1737.2-2858$, it is a power-law model with $\Gamma=2.49\pm 0.14$. Both sources are detected in the $\sim 0.3-10$~GeV energy range, and the yearly light curves are not consistent, although the error bars are so large to be consistent with a constant flux.

I have also downloaded the \emph{Fermi}/LAT data from Aug 5, 2008, 00:00:00 UTC to Mar 3, 2026, 23:59:57 UTC centered on 4FGL~J$1737.1-2901$ and within a radius of $15^{\circ}$. Then, I have analyzed those data by using \texttt{Fermi Tools v. 2.5.1} in the \texttt{conda} environment. I have selected the Instrument Response Function (IRF) \texttt{P8R3\_SOURCE\_V3}, the isotropic background model \texttt{iso\_P8R3\_SOURCE\_V3\_v1}, and the Galactic diffuse model \texttt{gll\_iem\_v07.fits}. Given the average spectra of the above cited catalogs (cf Fig.~\ref{fig:lat}), and given the strong diffuse emission due to the close Galactic centre, I have selected the \texttt{PowerLaw2} as source model, and the $0.3-10$~GeV energy band, where the source is detected. The energy band is also useful to minimize the issues due to the Galactic diffuse background. The $0.3-10$~GeV flux was then extrapolated to the $0.1-20$~GeV energy band by using the measured photon index to compare with \emph{CGRO}/EGRET measurements

The \texttt{xml} model file also includes all the sources (11) from the 4FGL within $1^{\circ}$ from 4FGL~J$1737.1-2901$. The 4FGL reported 286 gamma-ray sources within $10^{\circ}$ from 4FGL~J$1737.1-2901$, but all their models were not included into the \texttt{xml} file: this can affect the error estimates, but I needed to have reasonable computer times. Nonetheless, the results are consistent with the scenario emerged from the LAT catalogs. I did not include 4FGL~J$1737.1-2901$ in the \texttt{xml}, but GRS~$1734-292$: there is a difference of $\sim 8'$ between the coordinates of the two sources, and I wanted to test for emission specifically associated with with the Seyfert galaxy, and not the unassociated gamma-ray source of the 4FGL catalog. 

I ran two cycles of \texttt{gtlike}, one with the DRMNFB optimizer, and the other with NEWMINUIT, by using the output model from the previous cycle. 

I selected three time bins: one week around the X-ray observations, to search for simultaneous gamma-ray activity, four months, to search for some more detections on time scales shorter than the yearly light curve provided by the LAT catalogs, and one year, to check if GRS~$1734-292$ can really be linked to the \emph{Fermi} gamma-ray source in the 4FGL. If the data analysis resulted in zero photons, the upper limit at 95\% confidence level was estimated according to the LAT sensitivity for the selected exposure and assuming $\Gamma=2$, and selecting the worst case, given the high Galactic diffuse background. The most conservative values extrapolated to the $0.1-20$~GeV band are:

\begin{itemize}
\item 1 week exposure around the X-ray observation: $\sim 1.3\times 10^{-9}$~erg~cm$^{-2}$~s$^{-1}$; 
\item 4 months exposure: $\sim 1.3\times 10^{-10}$~erg~cm$^{-2}$~s$^{-1}$;
\item 1 year exposure: $\sim 8\times 10^{-11}$~erg~cm$^{-2}$~s$^{-1}$.
\end{itemize}

\begin{figure}[h!]
\begin{center}
\includegraphics[width=\textwidth]{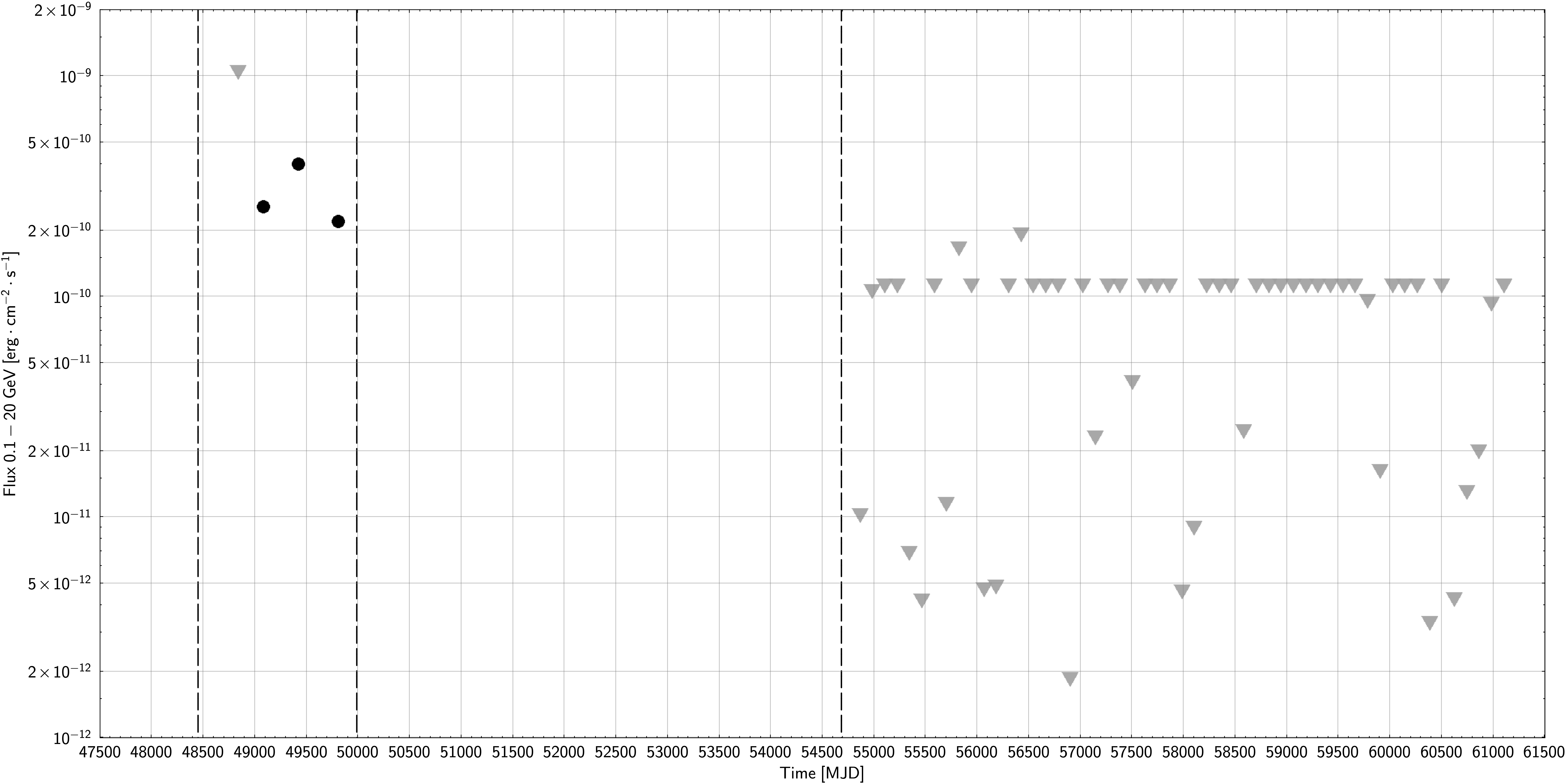}\\
\includegraphics[width=\textwidth]{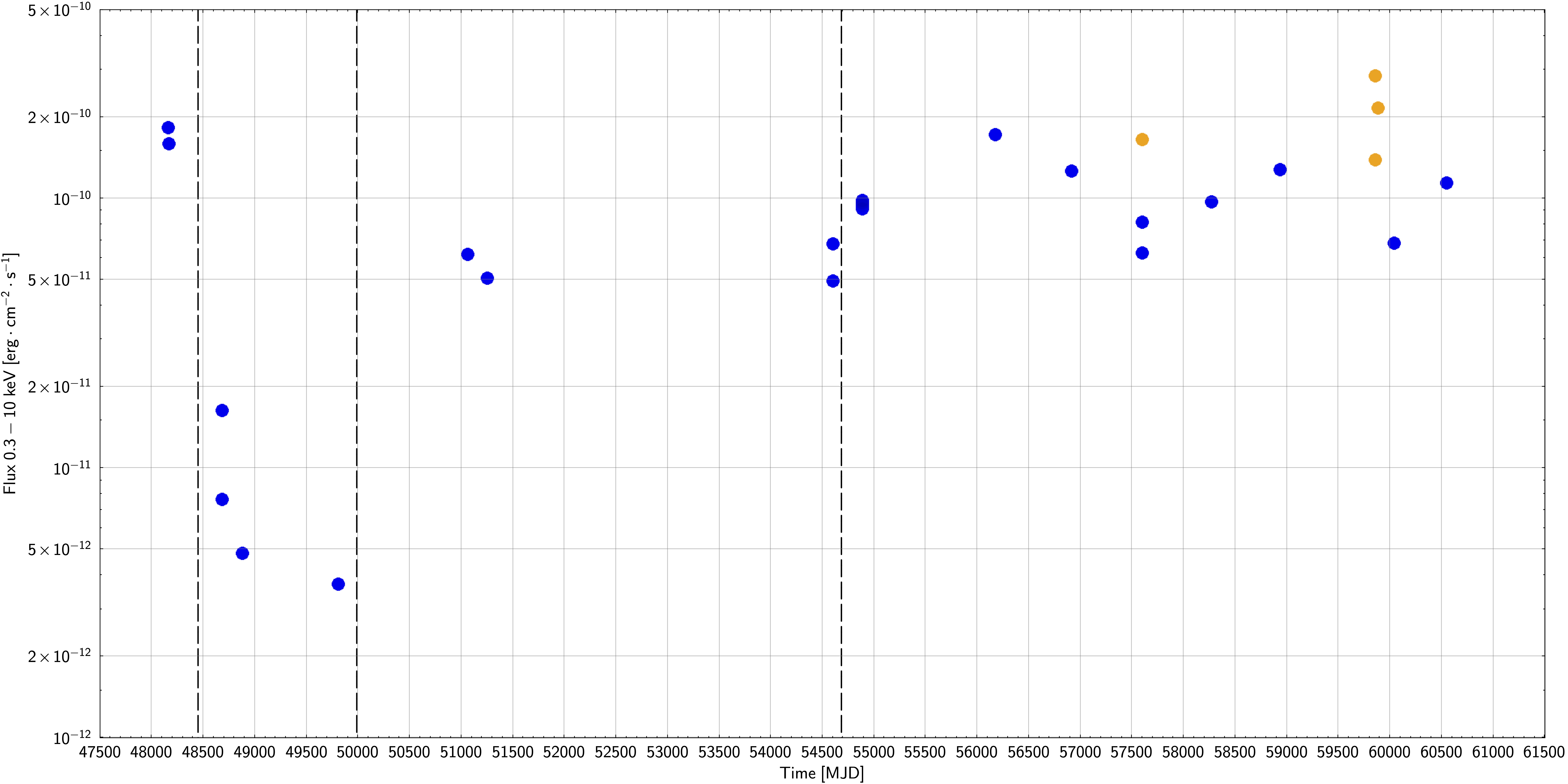}
\caption{(\emph{top panel}) Gamma-ray light curve in the $0.1-20$~GeV energy band. The dashed vertical lines represent the period covered also by \emph{CGRO}/EGRET observations (July 12, 1991 -- September 27, 1995, MJD $48449-49987$), and the beginning of scientific operations of \emph{Fermi}/LAT (Aug 5, 2008, MJD $54683$). EGRET points refer to the data integrated over the four cycles (P1, P2, P3, and P4, \citealt{HARTMAN1999}). The gray triangles indicate upper limits, while black circles are gamma-ray detections. \emph{Fermi}/LAT data refer to the 4-months time bins. (\emph{bottom panel}) X-ray light curve in the $0.3-10$~keV energy band. Blue circles refer to spectra modeled by using a power-law with intrinsic absorption, while orange circles are X-ray observations displaying absorption edges.}
\label{fig:xgammaflux}
\end{center}
\end{figure}

Fig.~\ref{fig:xgammaflux}, (\emph{top panel}), displays the 4-months light curve. The weekly exposures returned no detection, as did the 4-month bins, although, by considering the $2\sigma$ criterion adopted by \cite{HARTMAN1999}, there would be two cases, one with $TS\sim 9$, and the other with $TS\sim 7$. However, by using the threshold adopted by the \emph{Fermi}/LAT collaboration ($TS\geq 25$), there are no detections at all.

\section{Mass of the central black hole}
\label{mass}
There are two measurements of the mass of the central black hole of GRS~$1734-292$. \cite{TORTOSA2017} estimated the mass of the central black hole $M\sim 3.2\times 10^{8}M_{\odot}$ by using the H$\alpha$ emission line measured by the European Southern Observatory (ESO) New Technology Telescope (NTT) grism 13. 

\cite{KOSS2022} studied a sample of hard X-ray AGN selected by using \emph{Swift}/Burst Alert Telescope (BAT) and GRS~1734-292 is the source n. 896. It is classified as a Seyfert 1.9 (optical spectrum with narrow H$\beta$ and broad H$\alpha$) with $z=0.02176$, a measured mass of the central black hole $M\sim 6.9\times 10^{7}M_{\odot}$, accretion rate $\lambda_{\rm Edd}\sim 0.085$. \cite{KOSS2022} used ESO Very Large Telescope (VLT) with X-Shooter. In both cases, the error was evaluated to be $0.5$~dex, so the measurements are consistent within the errors, although the ESO/NTT has a lower resolving power at $\sim 6500$\AA\, ($\Delta \lambda/\lambda \sim 300$ vs 8900 of X-Shooter). 

As reference for calculations, I consider the arithmetic mean of $M\sim 1.5\times 10^{8}M_{\odot}$, corresponding to a gravitational radius $r_{\rm g}\sim 2.2\times 10^{13}$~cm. Given the optical spectra, it is clear a large viewing angle, which implies a classification as misaligned AGN, as also emerged from the radio observations by \cite{MARTI1998}, who detected a jet and a counterjet.

\begin{figure}[h]
\begin{center}
\includegraphics[width=\textwidth]{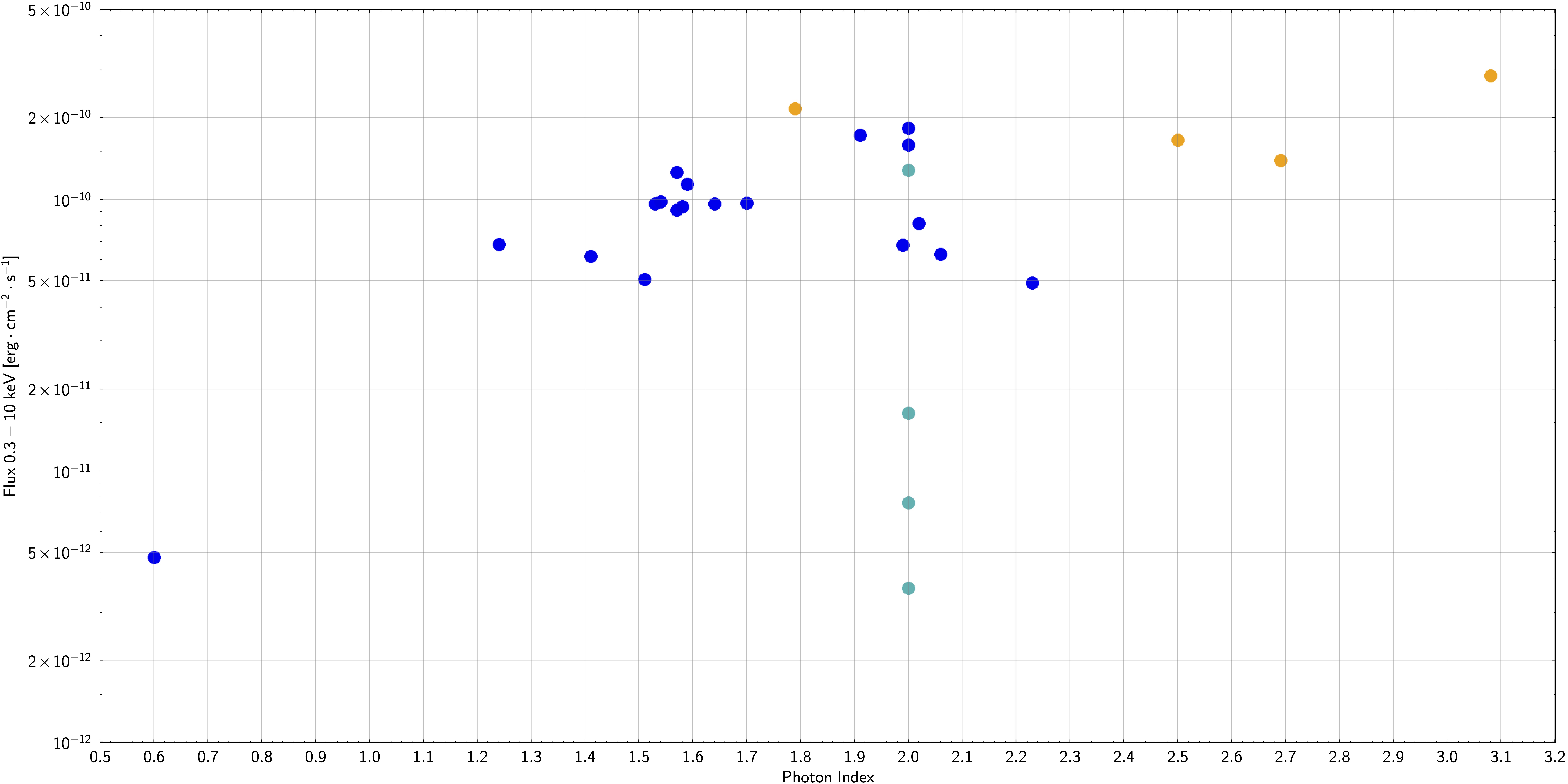}
\caption{X-ray light flux in the $0.3-10$~keV energy band vs photon index. Blue circles refer to spectra modeled by using a power-law with intrinsic absorption, while orange circles are X-ray observations displaying absorption edges. Gray circles indicate the use of a fixed $\Gamma=2$, because no measurement was found.}
\label{fig:index}
\end{center}
\end{figure}

\section{Overview and comparison with previous studies}
\label{overview}
Fig.~\ref{fig:xgammaflux} displays X- and gamma-ray fluxes as a function of time, including also historical observations. A first look shows that the gamma-ray detections with \emph{CGRO}/EGRET were recorded during very low X-ray fluxes in the $0.3-10$~keV energy band, while during the \emph{Fermi}/LAT observation, no gamma-ray detection was found in the periods simultaneous to the X-ray observations, but also in the 4-months time bins. X-ray observations during the \emph{Fermi}/LAT years measured higher fluxes, with evidence of absorption edges in some cases. 

The X-ray observation during the EGRET cycles were: 

\begin{itemize}
\item two observations with the Position Sensitive Proportional Counter (PSPC) onboard the \emph{ROSAT} satellite in February 1992 (see Table~\ref{tab:pastxray}); the flux nearly tripled in $\sim 4.8$ hours, from $\sim 0.03$ to $\sim 0.08$~c/s, with hardness ratio\footnote{Given the small energy band of PSPC ($0.1-2.4$~keV) and the strong absorption, both intrinsic and Galactic, the hard spectrum might simply be due to the absorption hampering the collection of photons, rather than an intrinsic hard photon index.} $1.00\pm 0.07$ \citep{VOGES1999};

\item the hard X-ray outburst on September 15-17, 1992, detected with \emph{GRANAT}/SIGMA, with $\sim 36$~mCrab in the $40-400$~keV energy band and a very hard photon index $0.6\pm 0.4$ \citep{CHURAZOV1992}; no detections were reported before and after the outburst, with $2\sigma$ upper limit $<20$~mCrab;

\item and the \emph{ROSAT}/HRI observation on Mar. 28-31, 1995 \citep{BARRET1996}, which is the lowest X-ray flux.
\end{itemize}

On the opposite, the lack of significant gamma-ray activity in more recent times is accompanied with higher X-ray fluxes, softer photon indexes, and the appearance of absorption edges (Fig.~\ref{fig:xgammaflux}, \ref{fig:index}). 

After the \emph{INTEGRAL} detection and the possible link with 3EG~J$1736-2908$ \citep{DICOCCO2004}, GRS~1734-292 gained renewed attention. \cite{SAZONOV2004} studied the X-ray spectrum in the $2-200$~keV energy band: in addition to the new data from \emph{INTEGRAL}, they retrieved old observations from \emph{ASCA} and \emph{GRANAT}/ART-P. The broadband spectrum is fit with a power law ($\Gamma\sim 1.7-1.9$), with intrinsic absorption ($N_{\rm H}^{z}\sim 2.0-2.2\times 10^{22}$~cm$^{-2}$), and with a cut-off at $E_{\rm cut}\sim 156$~keV. \cite{MOLINA2006} studied a sample of AGN on the Galactic plane, including GRS~$1734-292$, by using \emph{ASCA} and \emph{INTEGRAL} data. The resulting model is consistent with \cite{SAZONOV2004} ($N_{\rm H}^{z}\sim 1.05\times 10^{22}$, $\Gamma\sim 1.74$), except for the cut-off energy, which is quite smaller ($E_{\rm cut}\sim 58$~keV). A small cut-off energy is confirmed by \cite{MALIZIA2014} ($E_{\rm cut}\sim 58$~keV, with \emph{XMM-Newton}, \emph{INTEGRAL}, and \emph{Swift}/BAT), \cite{TORTOSA2017} ($E_{\rm cut}\sim 53$~keV, with \emph{XMM-Newton} and \emph{NuSTAR}), and \cite{RICCI2017} ($E_{\rm cut}\sim 84$~keV, with \emph{XMM-Newton}, and \emph{Swift}/BAT). 

The \emph{XMM-Newton} observation was first analyzed by \cite{GUAINAZZI2011}, who studied a sample of Seyfert to understand the relativistic effects on the features in the spectra of obscured objects (GREDOS sample, General Relativistic Effects Detected in Obscured Sources). His model is an absorbed power law ($N_{\rm H}^{z}\sim 1.41\times 10^{22}$, $\Gamma\sim 1.41$) with an upper limit for the equivalent width (EW) of the neutral Fe~K$\alpha$ line of $<420$~eV. A weak neutral Fe~K$\alpha$ ($EW=20\pm13$~eV) is instead reported by \cite{TORTOSA2017}, together with an absorption edge at $6.69$~keV corresponding to Fe~XXV~K shell ($EW=-31\pm 12$~eV), indicating the presence of a warm absorber. A more detailed study found the ionization parameter $\xi\sim 1778$~erg~cm~s$^{-1}$, column density $N_{\rm H}^{\rm wa}\sim 5.01\times 10^{22}$~cm$^{-2}$, and an upper limit to the velocity of the outflow $v_{\rm out}< 5300$~km~s$^{-1}$ \citep{TORTOSA2017}. 

More interesting is the cold cut-off energy, quite unusual when compared with other AGN. \cite{FABIAN2015} studied the coronal properties of a sample of AGN and black-hole binaries, focusing on the normalized corona temperature and the radiative compactness (see Fig.~1 in \citealt{FABIAN2015}). The first one is: 

\begin{equation}
\Theta = \frac{kT_{\rm e}}{m_{\rm e}c^2}
\label{eq:coronat}
\end{equation}

where $k$ is the Boltzmann constant, $T_{\rm e}$ is the electron temperature, $m_{\rm e}$ is the electron rest mass, and $c$ is the speed of light in vacuum, while the radiative compactness is:

\begin{equation}
\ell = \frac{L\sigma_{\rm T}}{R m_{\rm e} c^3}
\label{eq:compact}
\end{equation}

where $L$ is the source luminosity, $R$ is the source radius, $\sigma_{\rm T}$ is the Thomson cross section. \cite{TORTOSA2017} measured the coronal temperature $kT_{\rm e}\sim 12.1$~keV ($\Theta \sim 0.023$), and the optical depth $\tau\sim 2.8$ or $\sim 6.38$, depending on the geometry (slab or spherical, respectively). The radiative compactness $\ell\sim 13$ was calculated by assuming $R=10r_{\rm g}$, where $r_{\rm g}$ is the gravitational radius. These values place GRS~$1734-292$ close to the electron-proton coupling line in the Fig.~1 of \cite{FABIAN2015}. The high $\tau$ could explain the low temperature, as seed photons must scatter many times before exiting the corona, thus draining more energy from electrons \citep{TORTOSA2017}. 

It remains, however, the disagreement with the cut-off measured by \cite{SAZONOV2004}, who reported $E_{\rm cut}\gtrsim 110$~keV. Comparing with \cite{MOLINA2006}, who also used \emph{ASCA} archival data for the low-energy part of the spectrum, I noted these differences: (i) the IBIS exposure of \cite{SAZONOV2004} was $\sim 2$~Ms, while \cite{MOLINA2006} accumulated $\sim 4$~Ms of data; (ii) \cite{SAZONOV2004} used the archival \emph{ASCA} observation of March 1999, while \cite{MOLINA2006} used the September 1998 observation, but the spectral fits reported by \cite{SAKANO2002} are consistent within the measurement errors; (iii) \cite{SAZONOV2004} performed their own flux calibration of IBIS data by using Crab observations. It is also worth noting a study by \cite{MOLINA2013}, who analyzed a sample AGN detected both by \emph{INTEGRAL}/IBIS and \emph{Swift}/BAT. In the case of GRS~$1734-292$, they found a cut-off $E_{\rm cut}\gtrsim 160$~keV if the reflection is set to 1 and fixed, while if set to 0, then $E_{\rm cut}\sim 89$~keV. Indeed, \cite{SAZONOV2004} fixed the reflection to 1, while \cite{MOLINA2006} fixed it to zero. \cite{TORTOSA2017} measured $0.48\pm 0.22$ by using \emph{NuSTAR}, while \cite{RICCI2017} reported an upper limit of $0.8$.

The cold cutoff and the high optical depth are essential to explain the hard photon indexes (cf Fig.~\ref{fig:index}) as due to corona emission and not to a jet, with the exception of the \emph{GRANAT}/SIGMA outburst. 

\section{Variability}
\label{var}
The variability is the missing point in most of these studies, both on long (Fig.~\ref{fig:xgammaflux}) and short time scales (Table~\ref{tab:variability}). Some episodes of short term variability were observed. In addition to the \emph{ROSAT}/PSPC episode (see Sect.~\ref{overview}), on July 31, 2016, the flux nearly doubled in $\sim 0.75$~hr (from $8.15\times 10^{-11}$~erg~cm$^{-2}$~s$^{-1}$ to $1.65\times 10^{-10}$~erg~cm$^{-2}$~s$^{-1}$) and an absorption edge appeared at $\sim 1.68$~keV with $\tau\sim 0.74$ (Table~\ref{tab:chandra}). Another episode occurred on October 2, 2022: the flux increased from $1.39\times 10^{-10}$~erg~cm$^{-2}$~s$^{-1}$ to $2.85\times 10^{-10}$~erg~cm$^{-2}$~s$^{-1}$ in $\sim 0.92$~hr, and one more absorption edge appeared ($E_2\sim 2.62$~keV, $\tau\sim 0.41$, Table~\ref{tab:chandra}). The photon indexes are all consistent within the (large) measurement errors. 

\begin{table}[t]
\caption{Events of short variability. Columns: (1) Date of observation (cf Table~\ref{tab:pastxray}, \ref{tab:chandra}); (2) Time difference between the two observations [hours]; (3) Doubling/halving flux time linearly estimated [hours]; (4) Size of the emitting region [$r_{\rm g}$] calculated according Eq.~(\ref{eq:sizeer}).}
\begin{center}
\begin{tabular}{cccc}
\hline
Date & Observed $\Delta t$ & Doubling $\Delta t$ & Size\\
(1)  & (2)                & (3)          &  (4)\\
\hline
Feb 29, 1992 & 4.8 & 2.9 & 14\\
Jul 31, 2016 & 0.75 & 0.73 & 3.5\\
Oct 02, 2022 & 0.92 & 0.88 & 4.2\\
\hline
\end{tabular}
\end{center}
\label{tab:variability}
\end{table}

\begin{figure}[t]
\begin{center}
\includegraphics[width=\textwidth]{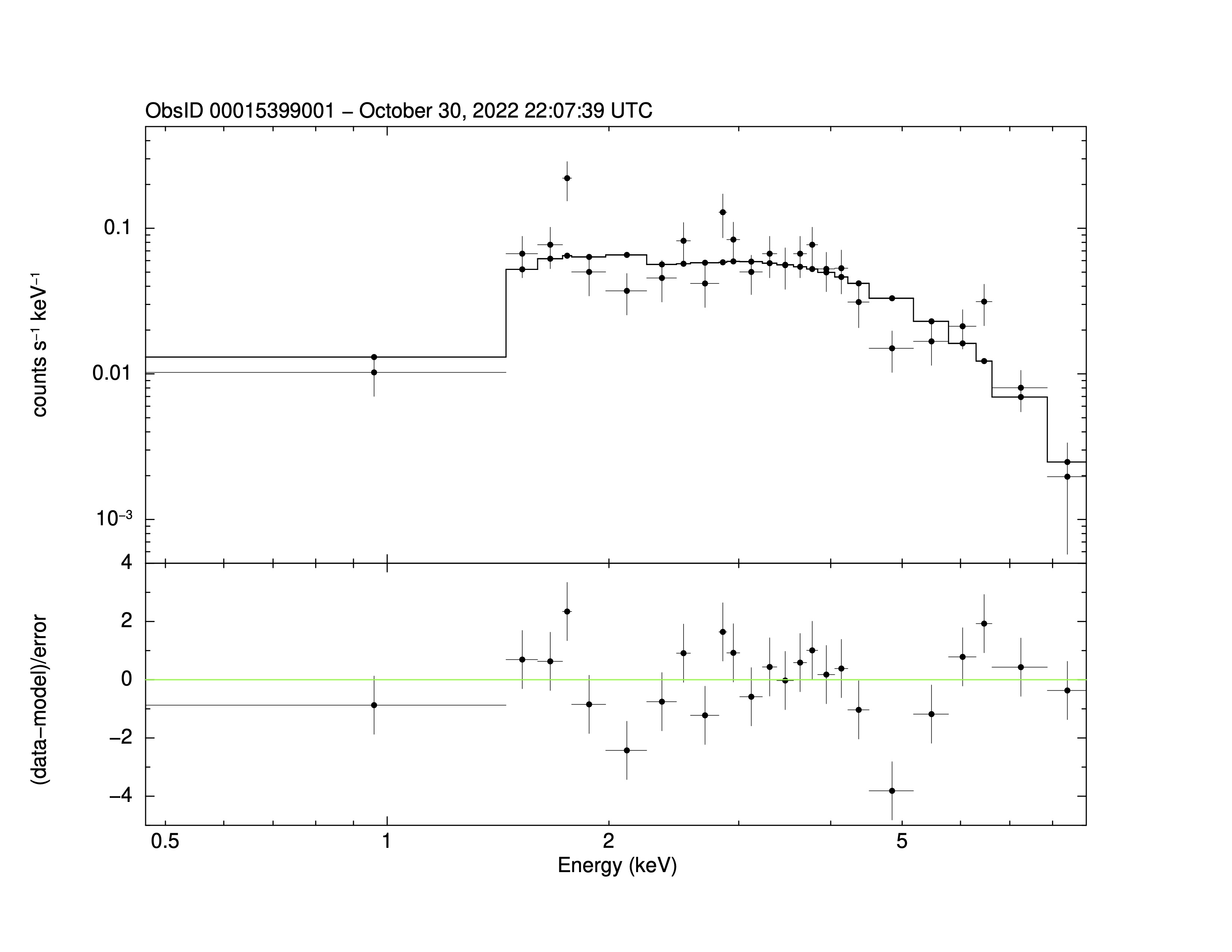}
\caption{\emph{Swift}/XRT observation id $00015399001$, on Oct. 30, 2022. The spectrum is fit only with the \texttt{xspec} model \texttt{tbabs*ztbabs*zpo} to emphasize the presence of absorption edges.}
\label{fig:edge}
\end{center}
\end{figure}

Let's start with the variations in absorption. Among the detected edges, three of them in the range $1.55-1.78$~keV could be associated with a Mg XI at $1.66$~keV, while two others in the ranges $1.86-2.05$~keV and $2.52-2.69$ could be due to silicon (Si XIII, $1.86$ keV; Si XIV, $2.00$ keV; Si XVI, $2.62$ keV). The edge at $4.25_{-0.17}^{+0.23}$~keV (detected on Oct 30, 2022, with \emph{Swift}) can be associated with Ca K edge: its value for neutral Ca is $4.04$~keV, but a slight ionization can push the edge at higher energies \citep{WITTHOEFT2009}. Such feature has been already detected in Cen~A \citep{MARKOWITZ2007}. However, the slightly higher energy might be also due to the poor statistics of the \emph{Swift} observation (Fig.~\ref{fig:edge}).  

The size of the edge region can be estimated as:

\begin{equation}
r < \frac{c\Delta t}{1+z}\sim 7.7\times 10^{13}~\mathrm{cm}\sim 3.5r_{\rm g}
\label{eq:sizeer}
\end{equation}

where $\Delta t$ is the minimum observed time scale of the doubling flux (Table~\ref{tab:variability}). By considering an outflow velocity $v_{\rm out}\lesssim 5300$~km~s$^{-1}$ (upper limit from \citealt{TORTOSA2017}), the region should be located at:

\begin{equation}
R > \frac{2GM}{v_{\rm out}^2}\sim 1.4\times 10^{17}~\mathrm{cm}\sim 6400r_{\rm g}
\label{eq:exter}
\end{equation}

corresponding to the border between the broad-line region (BLR) and the torus. Indeed, by converting the X-ray luminosity into bolometric luminosity by using the \cite{MARCONI2004} relationships, and using the \cite{BENTZ2013} relationship between the BLR size and the disk luminosity\footnote{\cite{BENTZ2013} used the $L_{5100\AA}$, so I calculated it from the bolometric luminosity by assuming the usual conversion $L_{5100\AA}\sim L_{\rm bol}/10$.}, then the size of the BLR is $\sim (4.2-8.3)\times 10^{3} r_{\rm g}$ (I have not considered the lowest X-ray flux, where it is possible a jet emission, as during the \emph{GRANAT}/SIGMA outburst; see below). Therefore, it is likely that the edges are caused by gas clouds in the broad-line region (BLR) passing through the line of sight, similar to other Seyfert galaxies (cf \citealt{BLUSTIN2005}).

The change of the X-ray continuum can be ascribed to the corona. Even though the photon indexes display no changes within the measurement errors, the increase in the X-ray flux suggest an injection of particles (changes of accretion), with almost constant electron temperature distribution and corona geometry. I have extrapolated the $0.3-10$~keV flux to $0.3-100$~keV band to have a better estimate of the corona luminosity, and calculated the compactness parameter of Eq.~(\ref{eq:compact}). The values of $\ell$ during the observation of July 31, 2016, were $45-64$, indicating a compact corona, while $\ell\sim 43-83$ in the case of the observation of October 2, 2022. 

To have a corona regulated by pair production, $\Theta < 0.2$ ($kT_{\rm e}<100$~keV, \citealt{FABIAN2015}), which in turn implies that a slab geometry is slightly preferred to explain the high compactness, the soft photon index and the small cut-off energy. 

In Feb 29, 1992, during \emph{ROSAT}/PSPC observations, the variability suggests a different corona (Table~\ref{tab:variability}): considering the observed fluxes (this time I extrapolated by assuming $\Gamma=2$, since there are no measurements of the photon index) and larger size of the emitting region ($14r_{\rm g}$), $\ell\sim 1-2$, consistent with the low flux and then, low accretion.  

On Sep 15-17, 1992, GRS~$1734-292$ underwent an outburst detected by \emph{GRANAT}/SIGMA: \cite{CHURAZOV1992} reported no detection on Sep 14, and 18, with $2\sigma$ upper limit $<20$~mCrab, while on Sep 15-17 the source was detected in the $40-400$~keV energy band with a rather hard photon index ($\Gamma\sim 0.6$) and flux $\sim 36$~mCrab, clearly indicating the absence of a cut-off. \cite{CHURAZOV1992} also set an upper limit for an emission line due to pair annihilation of $\sim 2\times 10^{-3}$~ph~cm$^{-2}$~s$^{-1}$ ($3\sigma$) in the $450-550$~keV energy band. These data were published with an IAU Circular, and I did not find any published paper on that episode, nor any indication of retraction in the following papers of the \emph{GRANAT} collaboration. \cite{BARRET1996} noted that GRS~$1734-292$ is not present in the integrated maps of the Galactic center made by \emph{GRANAT}/SIGMA: given the instrument sensitivity, the average flux should be less than $\sim 9$~mCrab in the $35-75$~keV energy band ($\sim 7.2\times 10^{-11}$~erg~cm$^{-2}$~s$^{-1}$), too high to constrain the corona. However, this outburst is crucial to brush up the relativistic jet hypothesis. The timing information provided by \cite{CHURAZOV1992} do not allow for a detailed calculation, so I consider one day as doubling time scale. The size of the emitting region according to Eq.~(\ref{eq:sizeer}) is then $\sim 115r_{\rm g}$. Extrapolating the flux during the outburst in the $0.3-100$~keV band, the compactness parameter is now $\ell\sim 1.3$, consistent with \emph{ROSAT} observations. 

It is interesting to add the gamma-ray observations into the problem. \emph{CGRO}/EGRET observations started in 1991, the early detections were found during the VP 27.0 (Apr 28 - May 7, 1992) and 210.0 (Feb 22-25, 1993), as shown in Fig.~\ref{fig:egret}. Please note that EGRET data in Fig.~\ref{fig:xgammaflux} referred to the integrated cycles, and P1 resulted in no detection, although there were some detections in the individual VPs as shown in Fig.~\ref{fig:egret}. There are no EGRET observations simultaneous or near the \emph{GRANAT}/SIGMA observation, but the two VPs before and after (27.0 and 210.0, see above) are detections with high flux (Fig.\ref{fig:egret}). I take as reference the weighted mean of the fluxes, which is $(1.17\pm 0.37)\times 10^{-6}$~ph~cm$^{-2}$~s$^{-1}$ in the $0.1-20$~GeV band, corresponding to $\sim 7.6\times 10^{-10}$~erg~cm$^{-2}$~s$^{-1}$, by using $\Gamma=2.18$, which is the average photon index reported in the Third EGRET Catalog \citep{HARTMAN1999}. The compactness is now $\ell\sim 8$. It is possible to estimate the optical depth for pair production in the EGRET energy band ($0.1-20$~GeV) by using Eqs.~(19) and (20) in \cite{MASTICHIADIS2002}\footnote{The definition of compactness in \cite{MASTICHIADIS2002} is different from that adopted by \cite{FABIAN2015} by a factor $4\pi$. This was taken into account in the calculations.}. From Eq.~(20), $\eta \sim 0.76$, and from Eq.~(19), $\tau_{\gamma\gamma}\sim 61-2350$ depending on the energy, and taking into account that the EGRET gamma-ray spectrum has no detection above $\sim 2$~GeV (see Fig.~\ref{fig:egret}, \emph{right panel}). To explain the gamma-ray detection, the optical depth has to be reduced by the relativistic beaming according to \citep{MASTICHIADIS2002}:

\begin{equation}
\delta > (\tau_{\gamma\gamma})^{\frac{1}{6}}
\label{eq:taubeam}
\end{equation}

where $\delta$ is the Doppler factor. With the above calculated values, it follows that $\delta \gtrsim 2-3.6$, which is not a particularly demanding value. In the following calculations, I will take as reference $\delta \gtrsim 3.6$, necessary to explain the detection of $\sim 2$~GeV photons. 

\section{Viewing angle}
Now, it is necessary to estimate the viewing angle, which is much more difficult. I found only one estimate by \cite{TORTOSA2017}, who adopted $\theta=60^{\circ}$ to fit the X-ray spectrum with the \texttt{pexrav} model for the Compton reflection. 

It is also possible to estimate the viewing angle from the optical spectrum: by using the flux table reported in Appendix of \cite{TORTOSA2017} and the methods reviewed by \cite{DALLABARBA2023}, taking into account the presence of the intrinsic absorption observed at X-rays (cf \citealt{DALLABARBA2024}), it is possible to infer $\theta \sim 40^{\circ}-60^{\circ}$. As an example, by using Eq.~(2) in \cite{DALLABARBA2023}:

\begin{equation}
\mathrm{Type}\sim 1+ \left[\frac{I(H\alpha_{\rm narrow})}{I(H\alpha_{\rm total})}\right]^{0.4}
\end{equation}

and taking the fluxes given by \cite{TORTOSA2017}, which are $I(H\alpha_{\rm narrow})\sim 3.0\times 10^{-15}$~erg~cm$^{-2}$~s$^{-1}$ and the $I(H\alpha_{\rm total})=I(H\alpha_{\rm narrow})+I(H\alpha_{\rm broad})\sim 2.2\times 10^{-14}$~erg~cm$^{-2}$~s$^{-1}$, then the Seyfert type is $\sim 1.45$ (intermediate). Another possibility is Eq.~(1) in \cite{DALLABARBA2023}:

\begin{equation}
R_{\rm opt}=\frac{I([OIII]5007)}{I(H\beta_{\rm broad})}
\end{equation}

but it can be applied only to the data from \cite{MARTI1998}, because \cite{TORTOSA2017} and \cite{OH2022} did not detect H$\beta$. From the data of \cite{MARTI1998}, it results $R_{\rm opt}\sim 2.4$, which confirms an intermediate classification. \cite{KOSS2022} classified GRS~$1734-292$ as Seyfert~$1.9$, but the H$\alpha$ flux published by \cite{OH2022} refer only to the total flux, and H$\beta$ is not detected, so it is not possible to estimate the viewing angle by using the above cited methods. 

It is also worth reminding the different epochs of optical observations: \cite{MARTI1998} and \cite{TORTOSA2017} used the ESO/NTT, the former observed in 1997, while the latter in 2010. In the 1997, the H$\beta$ is detected, while it is not in 2010, confirming a certain variability also at optical wavelengths. \cite{KOSS2022} and \cite{OH2022} observed at ESO/VLT on 2019 and confirmed no H$\beta$, while the total H$\alpha$ flux is lower than that measured by \cite{TORTOSA2017}: $\sim 1.5\times 10^{-15}$~erg~cm$^{-2}$~s$^{-1}$ vs $\sim 2.2\times 10^{-14}$~erg~cm$^{-2}$~s$^{-1}$. \cite{MARTI1998} reported $\sim 8.3\times 10^{-15}$~erg~cm$^{-2}$~s$^{-1}$. However, a certain caution is necessary given the uncertainties in the deblending of H$\alpha$ and [NII], as admitted by \cite{MARTI1998}, and the differences between the instruments.

Another method is from radio observations: \cite{MARTI1998} reported the presence of a jet extending $\sim 3''$ toward NE, plus a weak counterjet of $\sim 2''$. There are no flux measurements of these features, but one can estimate a flux ratio $R_{\rm j,cj}$ between the two features from the published maps in Fig.~1 of \cite{MARTI1998}. It resulted $R_{\rm j,cj}\sim 3.8$ and $R_{\rm j,cj}\sim 6.7$ for the $5$ and $8.6$~GHz map, respectively. I consider the latter as reference for calculation, as it should be less affected by opacity problems. Then, the viewing angle can be estimated by using (e.g. \citealt{GHISELLINI1993}):

\begin{equation}
R_{\rm j,cj}=\frac{S_{\rm j}}{S_{\rm cj}}\sim (\frac{1+\beta \cos \theta}{1-\beta \cos \theta})^{\epsilon+\alpha}
\label{eq:jetangle}
\end{equation}

where $\epsilon=2$ or $3$, depending on the jet type (continuous or discrete, respectively), $S_{\rm j}$ and $S_{\rm cj}$ are the flux densities of the jet and counterjet, respectively, $\alpha$ is the radio spectral index. \cite{MARTI1998} reported $\alpha=0.75$, and no indication of superluminal motion (the upper limit $5$~mas/day corresponds to $\beta_{\rm app}\lesssim 2600$, which is clearly not constraining), but the detection of the counterjet indicates no strong debeaming, and hence large viewing angles. The flux ratio sets an upper limit on $\theta\lesssim 71^{\circ}-76^{\circ}$, depending on the value of $\epsilon$. 

To obtain a viewing angle of $\theta \sim 60^{\circ}$ (cf \citealt{TORTOSA2017}), then $\beta\sim 0.50-0.66$, depending on $\epsilon$. According to the optical spectrum, GRS~$1734-292$ can be classified as intermediate. Therefore, by using $\theta \sim 45^{\circ}$, then $\beta\sim 0.35-0.47$. These values imply that the bulk Lorentz factor is $\Gamma_{\rm L}\sim 1.1-1.3$, and $\delta\sim 1.1-1.3$, which are clearly insufficient. 

However, if we take into account the absorbed jets (cf \citealt{LAHTEENMAKI2018,BERTON2020,JARVELA2021,ROMANO2023,JARVELA2024}), then $R_{\rm j,cj}$ would be altered and can increase with frequency and depending on the appearance or not of the absorber along the line of sight. The whole jet deprojected length can be estimated as $\sim 2.5-3.4$~kpc, depending on $\theta$ (given $1'' \ \mathrm{at}\ 90.2\,\mathrm{Mpc} = 0.437\,\mathrm{kpc}$). Therefore, given such extension, it is reasonable that there could be different absorption on the jet and the counterjet.

To obtain the required $\delta \sim 3.6$, then it would be necessary to have $R_{\rm j,cj}\sim 5800$ ($\epsilon=2$, continuous jet, which is the less demanding case), which in turn implies $\theta\sim 17^{\circ}$, $\beta_{\rm app}\sim 3.25$, $\beta\sim 0.96$, and $\Gamma_{\rm L}\sim 3.4$. The differential absorption should be:

\begin{equation}
R_{\rm j,cj}^{\rm obs}=R_{\rm j,cj}^{\rm int} e^{-(\tau_{\rm j}-\tau_{\rm cj})}
\label{eq:abs}
\end{equation}

where $R_{\rm j,cj}^{\rm obs}$ is the observed jet/counterjet flux ratio, $R_{\rm j,cj}^{\rm int}$ is the intrinsic one, $\tau_{\rm j}$ is the optical depth for the jet, and $\tau_{\rm cj}$ is that for the counterjet. Then, from Eq.~(\ref{eq:abs}), it results that $(\tau_{\rm j}-\tau_{\rm cj})\sim 6.8$ at $8.6$~GHz. Since the free-free absorption depends on the frequency as $\tau_{\rm ff}\propto \nu^{-2.1}$, it follows that at $37$~GHz (frequency of the Mets\"ahovi radio observatory), $\tau_{\rm ff,37GHz}\sim 0.32$, optically thin, as expected. Or, better, $\tau_{\rm ff}=1$ when $\nu\sim 21$~GHz. The fact that we still observe weak low-frequency radio emissions despite such high absorption could be explained by the relative proximity of GRS~$1734-292$. 

Obviously, it is not only a matter of absorption, but it is also necessary that the jet is in outburst, so to have an inverted radio spectrum (cf \citealt{JARVELA2024}). The mJy detections of GRS~$1734-292$ with ALMA \citep{MICHIYAMA2024} can be explained with a jet in low or off-state. 

\section{Gamma-ray emission: jet power vs winds}
I would like to compare the gamma-ray flux observed by \emph{CGRO}/EGRET with the jet and the wind power. I consider the gamma-ray flux around the \emph{GRANAT}/SIGMA outburst, which was $\sim 1.17\times 10^{-6}$~ph~cm$^{-2}$~s$^{-1}$ (cf Sect.~\ref{var}). By using the average photon index $\Gamma\sim 2.18$ observed by EGRET, I perform the $k-$correction ($z=0.02176$), conversion into cgs units and calculation of the luminosity, which results to be $L_{\gamma}\sim 7.4\times 10^{44}$~erg~s$^{-1}$. The radiative jet power can be estimated as (e.g. \citealt{GHISELLINI2014}):

\begin{equation}
P_{\rm rad}\sim 2\frac{\Gamma_{\rm L}^2}{\delta^4}L_{\gamma}
\label{eq:prad}
\end{equation}

\noindent By using the value of $\Gamma_{\rm L}\sim 3.4$ and $\delta\sim 3.6$ estimated in the previous section, the radiative jet power is $P_{\rm rad}\sim 1.0\times 10^{44}$~erg~s$^{-1}$. 

\begin{figure}[t]
\begin{center}
\includegraphics[width=\textwidth]{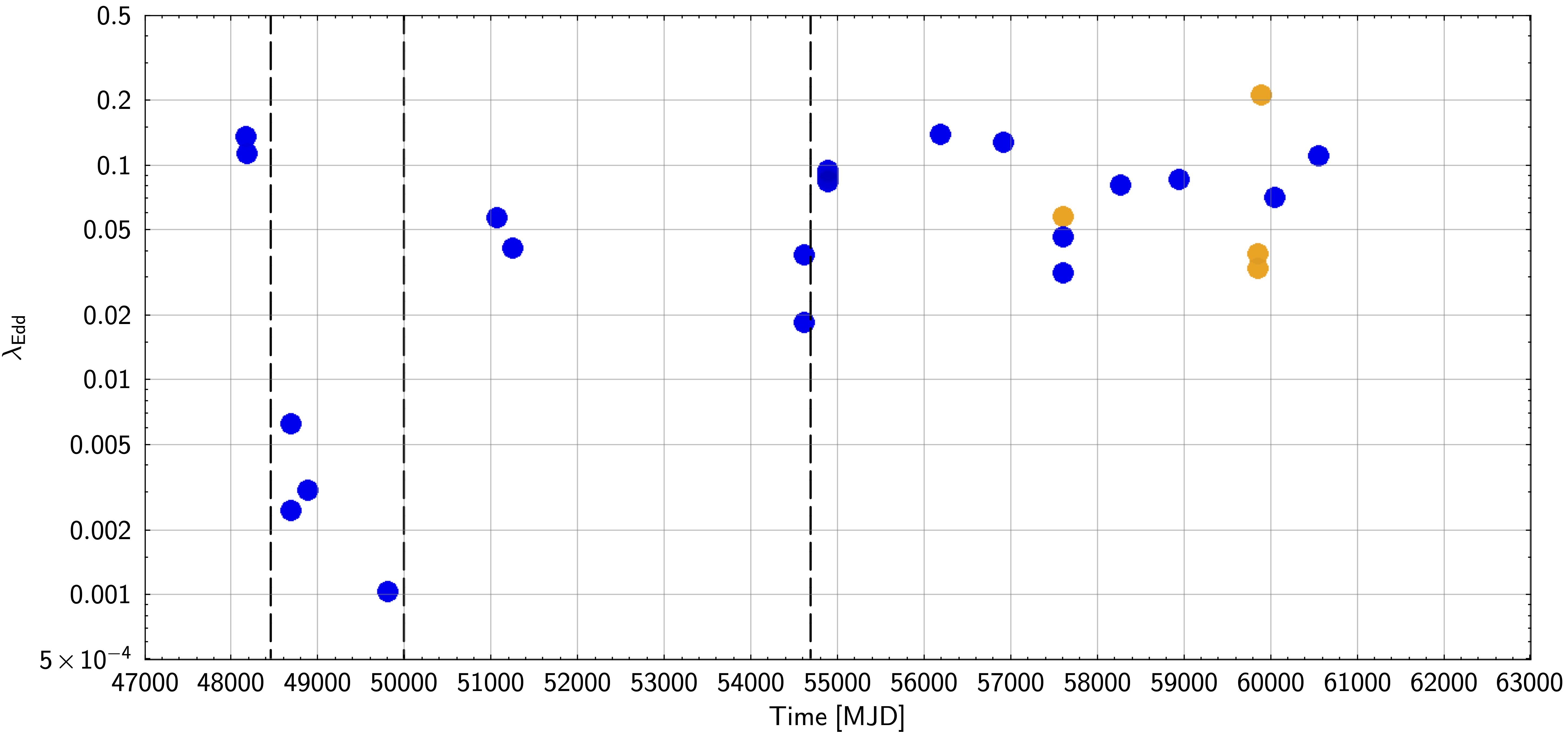}
\caption{Eddington ratio $\lambda_{\rm Edd}=L_{\rm bol}/L_{\rm Edd}$ vs time. X-ray spectra with absorption edges are shown with orange filled circles. The dashed vertical lines represent the period covered also by \emph{CGRO}/EGRET observations (July 12, 1991 -- September 27, 1995, MJD $48449-49987$), and the beginning of scientific operations of \emph{Fermi}/LAT (Aug 5, 2008, MJD $54683$). }
\label{fig:accretion}
\end{center}
\end{figure}

To compare with the Blandford-Znajek theory, it is necessary to know the type of accretion disk, because it can affect the magnetic field geometry \citep{GHOSH1997}. Therefore, I have converted the X-ray flux into the bolometric luminosity according to the relationships by \cite{MARCONI2004}, and measured the Eddington ratio ($\lambda_{\rm Edd}=L_{\rm bol}/L_{\rm Edd}$), which can be considered a proxy of the accretion rate $\dot{m}$. The results are displayed in Fig.~\ref{fig:accretion}. The \emph{GRANAT}/SIGMA outburst is at $\lambda_{\rm Edd}\sim 3\times 10^{-3}$, which suggests a gas-pressure dominated disk, confirmed also by the low value of $\ell\sim 1.3$, which indicates a diffuse corona. From Eq.~(14) of \cite{GHOSH1997}, the radiative power according to the Blandford-Znajek process is:

\begin{equation}
L_{\rm BZ}= 8\times 10^{42} M_{8}^{11/10} \dot{m}_{-4}^{4/5} a^2\; \mathrm{[erg~s^{-1}]}
\label{eq:bz}
\end{equation}

where $M_{8}$ is the black hole mass in units of $10^{8}M_{\odot}$, $\dot{m}_{-4}$ is the accretion rate in units of $10^{-4}\dot{m}$, and $a$ is the spin. In the present case, $M_{8}=1.5$ (cf Sect.~\ref{mass}), $a\sim 0.996$ (maximally rotating black hole), and $\dot{m}_{-4}=30$ if one takes $\lambda_{\rm Edd}\sim \dot{m}$, although when the disk is radiatively inefficient, the efficiency decreases, so the accretion rate might be greater than the Eddington ratio. By considering $\dot{m}=\sqrt{0.01\lambda_{\rm Edd}}$ \citep{NARAYAN2008}, then $\dot{m}_{-4}=56$. Therefore, the expected radiative power from Eq.~(\ref{eq:bz}) is $L_{\rm BZ}\sim (1.9-3)\times 10^{44}$~erg~s$^{-1}$, consistent with the EGRET observations. I have assumed a maximally rotating black hole, but it is possible to obtain a better agreement with a moderate spin $a\sim 0.6$. 

Relativistic jets are not the only structures able to generate gamma rays. \cite{LENAIN2010} found that the gamma-ray emission of the radio-weak Seyfert 2 galaxy NGC~1068 cannot be explained by the starburst phenomenon alone and that an additional component is required; they proposed that it arises from external inverse Compton scattering of infrared photons from the torus by a relativistic wind. Other authors \citep{LAMASTRA2016,AJELLO2021,INOUE2024,PERETTI2025} have refined this idea, introducing hadronic variants based on shocks and pion decay and that link these winds to cosmic rays and neutrinos. According to \cite{AJELLO2021}, these winds convert $\sim 0.04$\% of their kinetic power to gamma rays. 

Therefore, the next step is to estimate the kinetic power of the winds detected via the absorption edges (see Sect.~\ref{var}). The absorption edges at $1.66$~keV (Mg XI, $\tau\sim 0.51-0.75$) and $2.62$~keV (Si XVI, $\tau\sim 0.41$) can be used to estimate the ionic column density:

\begin{equation}
N_{\rm ion}=\frac{\tau}{\sigma(E)}
\end{equation}
 
where $\sigma(E)$ is the photon absorption cross section, whose values can be found in the tables by \cite{VERNER1996}. The values are $N_{\rm ion}\sim (1-4)\times 10^{18}$~cm$^{-2}$. The hydrogen column of the outflow $N_{\rm H}^{\rm out}$ can then be estimated:

\begin{equation}
N_{\rm H}^{\rm out}=\frac{N_{\rm ion}}{A_{\rm el} f_{\rm ion}}
\label{eq:hydrocol}
\end{equation}

where $A_{\rm el}$ is the abundance of the element with respect to hydrogen (I assume Solar abundance), and $f_{\rm ion}$ is the ionization fraction. \cite{TORTOSA2017} estimated the ionization parameter $\xi \sim 1778$~erg~cm~s$^{-1}$, which implies $f_{\rm ion}\sim 0.2-0.4$. Therefore, it results $N_{\rm H}^{\rm out}\sim (1-3)\times 10^{23}$~cm$^{-2}$, greater by a factor $\sim 2-6$ than the value $\sim 5\times 10^{22}$~cm$^{-2}$ found by \cite{TORTOSA2017}, by using the \emph{XMM-Newton} observation, while the present estimates were obtained from \emph{Chandra} and \emph{Swift} observations in other epochs. Given the large margins of error and the simplifying assumptions, the results can be considered consistent. 

The mass outflow rate $\dot{M}_{\rm out}$ can be estimated by using (e.g. \citealt{BLUSTIN2005}, by assuming $\Delta r/r\sim 1$):

\begin{equation}
\dot{M}_{\rm out}\sim 4\pi R N_{\rm H}^{\rm out} \mu m_{\rm p} v_{\rm out} C_{\rm f}
\label{eq:massout}
\end{equation}

where $R\sim 6400r_{\rm g}$ from Eq.~(\ref{eq:exter}), $m_{\rm p}$ is the proton rest mass, $\mu\sim 1.4$ is a parameter to take into account the presence of elements other than hydrogen, $v_{\rm out}\lesssim 5300$~km~s$^{-1}$ is the outflow velocity from \cite{TORTOSA2017}, and $C_{\rm f}\sim 0.1$ is the hypothesized covering fraction. By taking as reference $N_{\rm H}^{\rm out}= 2\times 10^{23}$~cm$^{-2}$, then $\dot{M}_{\rm out}\sim 4.4\times 10^{25}$~g~s$^{-1}\sim 0.70M_{\odot}$~yr$^{-1}$. The kinetic power is then:

\begin{equation}
L_{\rm kin}=\frac{1}{2}\dot{M}_{\rm out} v_{\rm out}^2\sim 6.2\times 10^{42}\, \mathrm{erg~s^{-1}}
\label{eq:ekin}
\end{equation}

Taking into account that only a fraction of the kinetic power is converted into gamma rays (cf \citealt{AJELLO2021}, $\eta\sim 4\times 10^{-4}$), then the expected gamma-ray flux is $F_{\gamma}\sim 2.6\times 10^{-15}$~erg~cm$^{-2}$~s$^{-1}$, well below the \emph{Fermi}/LAT sensitivity. Even by assuming a very optimistic value of $\eta\sim 0.01$ \citep{LAMASTRA2016}, the flux can increase only to $F_{\gamma}\sim 6.4\times 10^{-14}$~erg~cm$^{-2}$~s$^{-1}$, still outside the \emph{Fermi}/LAT performance. 

\cite{SAKAI2025} proposed a lepto-hadronic model, where an accretion disk wind interacted with the nearby environment and generated gamma rays via $pp$ interactions or external Compton emission. Their model is consistent with the flux of 4FGL~J$1737.1-2901$, but it requires an ultrafast outflow (UFO) with $v_{\rm UFO}\sim 0.3c$ and that a significant fraction ($f_{\rm UFO}=10-40$) of the radiation momentum to be converted into outflow momentum, neither of which has observational support. \cite{TORTOSA2017} reported a tentative detection with \emph{XMM-Newton}/EPIC-pn of a blueshifted Fe XXVI line at $\sim 9500$~km~s$^{-1}$ ($\sim 0.03c$), but the line is not confirmed by EPIC/MOSes. In addition, the requirement $f_{\rm UFO}=10-40$ becomes even more unfavorable if we add the measurements reported in this work, namely that the absorption edges have $\tau \lesssim 1$, with the exception of the Oct 30, 2022, where $\tau \sim 1-2$, and $\lambda_{\rm Edd}\lesssim 0.2$. With these values, one expect $f_{\rm UFO}\sim 1-2$ with $v_{\rm UFO}\sim 0.3c$ (cf \citealt{NOMURA2016,NOMURA2017}). 

These $pp$ interactions generated by these winds can decay into neutrinos $\pi^\pm \to \mu^\pm \to e^\pm + \nu$, which could escape easier than gamma rays, because there is no pair production. However, given the above estimates from Eq.~(\ref{eq:ekin}) and by assuming that the neutrino luminosity (all flavors) is of the same order of magnitude of the gamma-ray luminosity, the expected flux of neutrinos is $\sim (2-5)\times 10^{-15}$~erg~cm$^{-2}$~s$^{-1}$, well below the IceCube sensitivity \citep{ABBASI2025}.

\section{The proposed scenario}
From the collected observations, it is evident that GRS~$1734-292$ underwent a major change during these decades, likely due to a change in the accretion and obscuration. By looking at Fig.~\ref{fig:xgammaflux} and \ref{fig:accretion}, it is evident that the absorption edges appeared when $\lambda_{\rm Edd}\gtrsim 0.03$, while the gamma-ray activity occurred when $\lambda_{\rm Edd}\lesssim 0.01$ or even less, because the \emph{GRANAT}/SIGMA outburst is at $\lambda_{\rm Edd}\sim 3\times 10^{-3}$. This behavior is reminiscent of jets from X-ray binaries, which are launched when the source is in low/hard state, and suppressed in high/soft state. However, in the present case, the observed behavior seems to be linked to variable absorption and changes in the accretion rate: low X-ray flux, diffuse corona, no winds, no absorption, but jet; high X-ray flux, compact corona, winds, absorption edges, no jet.

The low X-ray flux, a diffuse corona, and the \emph{GRANAT}/SIGMA outburst suggested that GRS~$1734-292$ could really be the source detected by \emph{CGRO}/EGRET at gamma rays, under the hypothesis of a behavior similar to the absorbed jets observed by \cite{LAHTEENMAKI2018,BERTON2020,JARVELA2021,ROMANO2023,JARVELA2024}. Order-of-magnitude estimates of the Doppler factor and the viewing angle show that an absorbed relativistic jet can explain the observations. The required absorption is consistent with observations. 

Some difference must be noted: the sample studied by \cite{LAHTEENMAKI2018} was made of NLS1s only and the observations lasted about four years, while GRS~$1734-292$ is a broad-line Seyfert and I have collected data for about 36 years, although the jet activity seems to be confined to the four years of \emph{CGRO}/EGRET. NLS1s generated multiple and erratic radio outbursts during the four years of observations, while this case presented only one episode attributable to a relativistic jet. On one side, this difference might be due to the Seyfert stage: NLS1s are young AGN, while GRS~$1734-292$ is likely older; on the other side, there was no monitoring of GRS~$1734-292$ during nineties, and the few observations were linked to the discovery and the 1992 \emph{GRANAT}/SIGMA outburst. 

Perhaps, GRS~$1734-292$ might be more similar to the case of NGC~4151, which underwent a giant gamma-ray outburst in late seventies \citep{DICOCCO1977} and then nothing more, although there are recent claims of gamma-ray emission from outflows (\citealt{PERETTI2025}; see \citealt{FOSCHINI2026} for a recent summary of the case). The similarity might play on these pivotal points: jet activity to explain the gamma-ray detection of NGC~4151 in late seventies, and of GRS~1734-292 in early nineties; winds to explain the recent activity. The \emph{Fermi}/LAT gamma-ray detection of NGC~4151 could be explained with more powerful winds and smaller distance with respect to GRS~1734-292, which was not detected at gamma rays. However, X-ray observations showed a change during the \emph{Fermi}/LAT years with respect to \emph{CGRO}/EGRET epoch. Some absorption edges appeared when the X-ray flux exceeds $\sim 10^{-10}$~erg~cm$^{-2}$~s$^{-1}$ or $\lambda_{\rm Edd}\sim 0.03$. The X-ray variability shows that the corona has become more compact, which in turn could imply a different geometry of the magnetic fields and, consequently, the suppression of the jet activity. Outflows appeared, although the estimation of the gamma-ray flux expected from the winds led to the conclusion that GRS~$1734-292$ cannot be detected by \emph{Fermi}/LAT (too far, too weak). The lack of significant gamma-ray activity in the recent years could also explain why the position of the gamma-ray source detected by \emph{Fermi}/LAT drifted away from GRS~$1734-292$ in the latest FL16Y catalog. The analysis of LAT data with different time scales (1 week, 4 months, 1 year) performed in this work, specifically looking for gamma-rays from the position of GRS~$1734-292$, confirmed that the Seyfert cannot be the counterpart of the \emph{Fermi}/LAT gamma-ray source. 

All this is circumstantial evidence, the real smoking gun is missing, which would be the observation of strong high-frequency ($\nu\gtrsim 21$~GHz) radio outbursts, such as those observed by \cite{LAHTEENMAKI2018} and \cite{JARVELA2024}. It is not clear if starting a radio monitoring now could lead to some outburst detection: it is evident that GRS~1734-292 changed its structure and activity, but it is unknown if this is a permanent change or not, and if there is still the possibility of a jet development. Since GRS~$1734-292$ is clearly a more evolved object than NLS1s, it is possible that the gamma-ray outburst observed by EGRET was the last gasp of a source entering an old age phase. The only way to discover it is to monitor GRS~$1734-292$ with high-frequency radio observations.

\section*{Acknowledgements}
This research has made use of data and/or software provided by the High Energy Astrophysics Science Archive Research Center (HEASARC), which is a service of the Astrophysics Science Division at NASA/GSFC. 

This research has made use of Aladin sky atlas developed at CDS, Strasbourg Observatory, France \citep{BAUMANN2022}.

\end{document}